\documentclass[12pt]{article}

\usepackage{newtxtext,newtxmath,accents}

\usepackage{graphicx}

\usepackage[letterpaper,margin=1in]{geometry}

\renewenvironment{abstract}
	{\quotation}
	{\endquotation}

\date{}

\makeatletter
\renewcommand{\fnum@figure}{\textbf{Figure \thefigure}}
\renewcommand{\fnum@table}{\textbf{Table \thetable}}
\makeatother

\usepackage{scicite}

\usepackage{url}

\RequirePackage{mathtools,amsmath,amsfonts,xspace} 

\usepackage{arydshln}

\newcommand{\totem}{\textsc{Tot}\textsc{Em}\xspace} 

\DeclareMathOperator*{\argmin}{argmin}

\DeclareMathOperator*{\rank}{rank}
\DeclareMathOperator*{\D}{D}

\DeclarePairedDelimiterX{\infdivx}[2]{(}{)}{%
  #1\;\delimsize\|\;#2%
}
\newcommand{\infdiv}{\D\infdivx}

\def\scititle{
	A totally empirical basis of science
}
\title{\bfseries \boldmath \scititle}

\author
{Orestis Loukas,$^{1}$ Ho-Ryun Chung$^{1\ast}$\\
\\
\normalsize{$^{1}$Institute for Medical Bioinformatics and Biostatistics, Philipps University Marburg,}\\
\normalsize{35043 Marburg, Germany}
\\
\normalsize{$^\ast$To whom correspondence should be addressed; E-mail:  ho.chung@staff.uni-marburg.de.}
}
\author{
	Orestis Loukas$^{1}$, Ho-Ryun Chung$^{1\ast}$\and
	\small{$^{1}$Institute for Medical Bioinformatics and Biostatistics, Philipps University Marburg, 35043, Germany}\and
    \small{$^\ast$To whom correspondence should be addressed; E-mail:  ho.chung@staff.uni-marburg.de.}
}

\begin{document} 

\maketitle

\begin{abstract} \bfseries \boldmath
Statistical hypothesis testing is the central method to demarcate scientific theories in both exploratory and inferential analyses. However, whether this method befits such purpose remains a matter of debate. Established approaches to  hypothesis testing make several assumptions on the data generation process beyond the scientific theory. Most of these assumptions not only remain unmet in realistic datasets, but often introduce unwarranted bias in the analysis. Here, we depart from such restrictive assumptions to propose an alternative framework of \textsc{tot}al \textsc{em}piricism.  We derive the Information-test ($I$-test) which allows for testing  versatile hypotheses including non-null effects. To exemplify the adaptability of the $I$-test to application and study design, we revisit the hypothesis of  interspecific metabolic scaling in mammals, ultimately rejecting both competing theories of pure allometry.
\end{abstract}

\noindent
The ability of theories to explain perceived phenomena is at the heart of science. Exploratory analysis aims to identify among many competing candidates a theory that best explains a phenomenon, whereas inferential analysis seeks to generalize a theoretical explanation from particular observations to the population. Common to both is the need to demarcate ``good'' from ``bad'' theories. Statistical hypothesis testing is the central method for such a demarcation. However, whether statistical hypothesis testing is suitable for this task is a matter of an ongoing debate~\cite{wasserstein_asa_2016,wasserstein_moving_2019,ioannidis_importance_2019,mayo_statistical_2022}. 
 
Critics are concerned about the adherence to null hypotheses with zero effect~\cite{meehl_theoretical_1978,mcshane_abandon_2019}, the validity of assumptions about the data generation process~\cite{mcshane_abandon_2019}, and whether the results of statistical hypothesis testing provide evidence for or against a hypothesis~\cite{berkson_difficulties_1938,edwards_bayesian_1963,berger_testing_1987,cohen_earth_1994,hubbard_why_2008}. Hypothesis testing procedures of the ``old statistics'' operate on a so-called test statistic and its distribution. The test statistic is computed from the manifestation of attributes in the data, while its distribution relies on strong assumptions about the data generation process. For example, Student's $t$-test for the equality of means in two sub-populations assumes that the data is normally distributed in both with equal variances. The $p$-values calculated from the $t$-distribution are only meaningful, if these assumptions are met. If this is not the case, a small $p$-value could signify that the means are not equal, but also that the assumptions about the data generation process are violated. Thus, the validity and inferential meaning of $p$-values become dependent on the validity of additional, theory-unrelated assumptions fueling the ongoing criticism of statistical hypothesis testing~\cite{nuzzo_scientific_2014}.

Apart from the criticism of statistical hypothesis testing \emph{per se}, there is a growing realization of an inflation of false positive findings due to the flexibility in data collection and analysis referred to as ``$p$-hacking'' \cite{simmons_false-positive_2011}. A solution to this problem consists in the prespecification of data collection and analysis \cite{ioannidis_importance_2019}. Different study designs, e.g.\ observational versus experimental, provide different prespecifiable information that should have repercussions to the statistical power of the test. Besides Pearson's, Barnard's and Fisher's test for contingency tables, the ``old statistics'' does not generically provide intrinsic means to account for the effect of the study design.

In this work, we do not aim at tackling all these problems within the ``old statistics''. Instead, we propose an alternative statistical framework which dispenses with assumptions on the data generation process, altogether. Relying on the universality of sampling in the statistical regime, i.e.\ at large sample size, our framework solely requires two things: the data and a hypothesis -- so we refer to it as \textsc{tot}ally \textsc{em}pirical (\totem) statistics. Combining well-known facts from information~\cite{csiszar_i-divergence_1975,csiszar_sanov_1984,cover_elements_2006} and probability~\cite{caticha_entropy_2021,Rosenkrantz1989} theory, which become relevant at large sample size, we derive a single statistical hypothesis test, the Information-test ($I$-test). The latter aims at replacing the multitude of statistical tests in the literature.

Taking into account the study design (i.e.\ how the data were collected by the researcher), \totem constructs the unique statistical population that minimally differs from the data but satisfies the hypothesis of the scientific theory. In such an ideal statistical population, we can efficiently approximate the probability ($p$-value) to observe a violation of the hypothesis at least as improbable as in the data. After revisiting the epistemological definition of an effective law, we demonstrate the applicability of our holistic framework by testing Kleiber's law \cite{kleiber_body_1932} -- an empirical power law that relates the basal metabolic rate (\textsc{bmr}) to the body mass of species.

\subsection*{Total Empiricism} 
Even when the scientific theory makes some predictions on the investigated matter, the specifics of data generation remain usually unknown or are intractable. This renders assumptions on some ``true'' data generation process unjustifiable and misleading. At best, sampled data provides the only reliable source of information about a population.

\paragraph*{On distributions}
A well-defined dataset is a collection of $N$ entries (complete rows in a table). At large $N$, even non-independently sampled data can approximately appear \cite{csiszar_sanov_1984} as independently sampled, which is our only working assumption. In scientifically interesting datasets, entries usually consist of manifestations of categorical (nominal or ordinal) and/or metric attributes, to which we refer by random variables $X,Y$ (generalization to more attributes being straight-forward). As random variables can neither continuously manifest nor extend over unbounded domains in an actual dataset, we concentrate on the finite set $\mathcal M$ of joint manifestations $(x,y)$ that actually appear in the data (see Table~\ref{tab:notation} for an overview).

Instead of the joint distribution of attributes, we investigate the distribution $\mathfrak p$ of joint manifestations of attributes, which assigns a probability $p(x,y)$ to each manifestation $(x,y)\in\mathcal M$. All these distributions are members of the probability simplex over $\mathcal M$. Colloquially, we refer to a dataset via the associated distribution whose probabilities are given by the relative frequencies of manifestations. In particular, the empirical distribution $\mathfrak f$ summarizes all the information about the population provided by the data.  

\paragraph*{From theory to conditions}
A scientific theory could posit a fundamental law such as a relationship $g(X,Y)$ between observable attributes. Any observed deviation would be traceable to measurement errors and/or some known source of systematic distortion. To actually prove the relationship one would need to assay the complete population. Since this is rarely possible or even well-defined, only a smaller sample is taken from the population. Beyond the theoretical relationship with its anticipated errors, the remaining properties of the sample --\,often needed in error estimation\,-- that allude to characteristics of the population must be estimated from the data. Testing scientific theories thus becomes a statistical problem.
 
In light of the inevitable measurement errors as well as some potential intrinsic variability and/or incompleteness of the theory, it often makes little sense to demand that every single observation $(x,y)$ systematically follows $g(x,y)$. Even if we cannot theoretically justify observed deviations from the conjectured relationship, we might at least anticipate  that the population obeys the relationship ``on average''. This yields an effective law as a condition on the expectation of $g$ in the population:
\begin{equation}
    \label{eqn::expectation}
    \langle g(X,Y) \rangle \overset{!}{=} \mu,
\end{equation}
where $\mu\in\mathbb R$ reflects the theoretical prediction. A perceived fundamental law like general relativity can give effective predictions like the average travel time of light from sun to earth,  but the converse cannot evidently hold. For example, an effective law of body mass predicting the average weight of \textsc{usa} population implies  neither that every citizen in the country shares the same value given similar anthropometric traits nor that all sub-population effects have been systematically accounted for. 

A statistical test needs to evaluate the probability to observe the empirical estimate of the expected relationship, i.e.\ the sample average $\langle g(X,Y)\rangle_\mathfrak{f}$ as if it had been sampled from a population obeying Equation~\ref{eqn::expectation}. 
To this end, conventional approaches involve the construction of a test statistic whose ideal distribution under Equation~\ref{eqn::expectation} is determined by assuming some data generation process. As the latter cannot be usually motivated by some fundamental scientific theory, we wish to avoid such assumptions altogether. Instead, we think along the lines of a most generic scenario where the condition in Equation~\ref{eqn::expectation} is precisely realized without any random-effect distortion. 

Starting from some anticipated relationships $g_a$ for $a=1,\ldots,D$, we want to construct an idealized population described by distribution $\mathfrak p : \mathcal M \rightarrow [0,1]^{\vert\mathcal M\vert}$ such that expectations 
\begin{equation}
    \label{eqn::TOTEMcondition}
    \langle g_a(X,Y) \rangle_\mathfrak{p} \equiv \sum_{(x,y)\in\mathcal M} g_a(x,y)\:p(x,y)\overset{!}{=}    \mu_a  ~,
\end{equation}
attain the desired values $\mu_a\in\mathbb R$ predicted by the theory. Conceptually, we distinguish between hypothesis and structural conditions: hypothesis conditions are directly deducible from theory, while structural ones correspond to expectations that need to be known and fixed.
This already preludes that certain (statistical) properties of the sample should be known before the analysis, i.e.\ should be prespecified. Trivially, normalization on the simplex with $g_\text{N}(x,y)=1$ and $\mu_\text{N}=1$ is always implied by the structural conditions.

For computational purposes, the distribution $\mathfrak p$ can be represented through a column vector that is indexed by manifestations $(x,y)\in\mathcal M$.
Similarly, functions $g_a$ can be thought of as  row-vectors that we summarize in a coefficient matrix $\mathbf{G}$. In that way, the $a$-th row of $\mathbf G$ describes the expectation Equation~\ref{eqn::TOTEMcondition} via a scalar product in $\mathbb R^{\vert\mathcal M\vert}$. In this representation, both hypothesis and structural conditions together form a system of equations linear in the probabilities of manifestations: 
\begin{equation}
     \label{eqn::HypothesisLinearSystem}
     \mathbf{G}\: \mathfrak{p} = \boldsymbol{\mu}_\text{H}
     \quad\text{where}\quad
     \mathfrak p\in\mathbb R^{\vert\mathcal M\vert}_{\geq0} 
     \quad\text{and}\quad \boldsymbol{\mu}_\text{H} \in\mathbb R^D
     ~.
\end{equation}
 
\paragraph{From conditions to the hypothesis distribution}
All distributions that constitute solutions of Equation~\ref{eqn::HypothesisLinearSystem} form a subspace on the probability simplex, referred to as the hypothesis \totem{plex} $\mathcal{T}_\text{H}$. 
In absence of further information, the data represented by $\mathfrak{f}$ corresponds to the most objective statistical model about the population, even if  generically $\mathfrak f\not\in\mathcal T_\text{H}$. 

Consequently, all we can do is to sample from $\mathfrak{f}$ datasets with $\mathfrak p\in\mathcal T_\text{H}$ of sufficiently large size that satisfy the hypothesized conditions. According to the conditional limit theorem~\cite{cover_elements_2006}, there is one dataset which is most likely to be sampled with its probability becoming one in the asymptotic limit of infinite sample size. The corresponding distribution, the so-called information projection ($I$-projection) of $\mathfrak{f}$ on $\mathcal{T}_\text{H}$, minimizes the information divergence ($I$-divergence; also called Kullback–Leibler divergence or relative entropy~\cite{kullback_information_1951}) from the data under the theoretical conditions. Most importantly, the $I$-projection always exists and is unique, if the \totem{plex} $\mathcal{T}_\text{H}$ is not empty~\cite{csiszar_i-divergence_1975}. Incidentally, the probabilities of the $I$-projection reflect the expected relative frequencies of manifestations over all datasets in the conditional sampling from $\mathfrak f$.
 
To avoid introducing bias warranted neither by theory nor data, we select the $I$-projection as the hypothesis distribution $\mathfrak{p}_\text{H}=\argmin_{\mathfrak p\in\mathcal T_\text{H}}\infdiv{\mathfrak p}{\mathfrak f}$. 
In that way, our hypothesis distribution describes an idealized population that is most similar to the original data, while \emph{exactly} fulfilling the hypothesis and structural conditions in Equation~\ref{eqn::HypothesisLinearSystem}. 
These considerations highlight a key difference of the \totem approach to the ``old statistics'': 
Instead of the idealized distribution of a test statistic under the conditions of theory and data generation, \totem obtains the manifestation distribution in an idealized population. The latter fulfills the conditions imposed by theory while incorporating any other theory-compatible effect from the data.

For the computation of the test statistic, conventional methods require auxiliary application-dependent criteria (such as the sample variance in the $t$-test), which correspond to additional conditions that unnecessarily obscure the analysis. In contrast, \totem solely reweighs the probabilities of manifestations via the minimization of the $I$-divergence to deduce the hypothesis distribution.
To accomplish this, the $I$-projection $\mathfrak{p}_\text{H}$ does not usually have a closed-form expression in terms of the values $\boldsymbol \mu_\text{H}$. 
Nevertheless, $\mathfrak{p}_\text{H}$ can be efficiently estimated by iterative algorithms \cite{csiszar_i-divergence_1975}, such as the iterative proportional fitting algorithm \cite{kruithof_telefoonverkeersrekening_1937,deming_least_1940} for marginal conditions and our Information-projector ($I$-projector; Supplementary Text - Background) for any type of linear conditions. 

\paragraph*{From the hypothesis distribution to the $p$-value}
To define a $p\text{-value}$ we formally need to change our perspective away from $\mathfrak f$ and sample from the hypothesis distribution.
Irrespective of the actual quasi-independent sampling process involved in data generation~\cite{csiszar_sanov_1984}, $\mathfrak p_\text{H}$ universally parametrizes at large sample size $N$ a multinomial sampling distribution (Supplementary Text - Background). This determines the probability of the data and, in fact, of any dataset of size $N$ under the hypothesis.

A $p$-value could be determined by summing the sampling probability under $\mathfrak p_\text{H}$ of  datasets which show the same or larger violation of  conditions in Equation~\ref{eqn::HypothesisLinearSystem} than the actual data. Albeit exactly calculable for smaller problems, identifying all possible datasets over manifestations $\mathcal M$ of size $N$ quickly becomes far from trivial, because their number exponentially increases with $N$ and $\vert\mathcal{M}\vert$. Alternatively, we could adopt the bootstrap~\cite{efron_introduction_1994} by sampling from the hypothesis distribution to approximate the $p$-value, which might be computationally expensive depending on the required accuracy. In fact, the suggested hypothesis distribution unambiguously solves the main problem of the bootstrap: sampling under the ``null hypothesis'' without resorting to arbitrary transformations of the data, see Supplementary Text - Application. 

Instead of bootstrapping, we can approximate the aforementioned cumulative sampling probability via a $\chi^2$ distribution. Datasets $\mathfrak p$ sampled from $\mathfrak p_\text{H}\in\mathcal T_\text{H}$ that satisfy the structural conditions, i.e.\ $\mathfrak p\in\mathcal T_\text{S}$ (the structural \totem{plex}), do not need to fulfill the hypothesis conditions ($\mathfrak p\not\in\mathcal T_\text{H}$). For each deviation from hypothesis conditions, there is a \totem{plex} consisting of all distributions exhibiting this numerical mismatch while fulfilling the rest of structural conditions. 
In particular, the alternative \totem{plex} 
$ \mathcal{T}_\text{A} = 
 \left\{\mathfrak{p} \,\vert\, \mathbf{G}\left(\mathfrak{p} - \mathfrak{f}\right) = 0 \right\}
$
encompasses all distributions that share the same expectations $\boldsymbol{\mu}_\text{A}\equiv \mathbf G\:\mathfrak f$ with the empirical distribution $\mathfrak{f}$. 

\begin{figure} 
	\centering
	\includegraphics[width=1\textwidth]{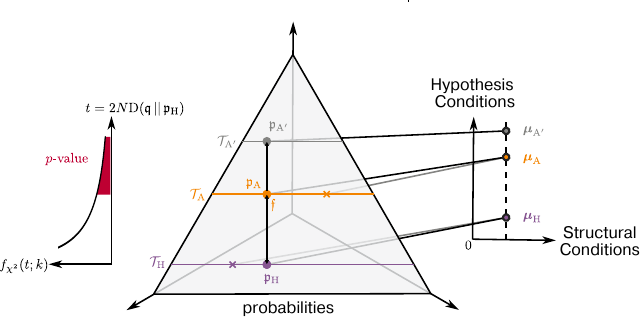} 

	\caption{\textbf{Total Empiricism.}
	    Probability simplex for three manifestations, where the only structural condition is the normalization. Starting from the data $\mathfrak{f}$ (orange dot), the hypothesis distribution $\mathfrak{p}_\text{H}$ (purple dot) is obtained by the $I$-projection of $\mathfrak{f}$ on the hypothesis \totem{plex} $\mathcal{T}_\text{H}$ (purple line). The hypothesis distribution $\mathfrak{p}_\text{H}$ as any other distribution $\mathfrak{q} \in \mathcal{T}_\text{H}$ (purple cross) fulfills the hypothesis conditions $\mathbf{G}\mathfrak{q} = \boldsymbol{\mu}_\text{H}$ (purple dot on the right). The probability of observing a deviation from the hypothesis as in the data is governed by the $I$-projection of $\mathfrak{p}_\text{H}$ on the alternative \totem{plex} $\mathcal{T}_\text{A}$ (orange line), which often coincides with the data $\mathfrak{p}_\text{A}=\mathfrak{f}$. Every member $\mathfrak{q}$ (orange cross) of the alternative \totem{plex} $\mathcal{T}_\text{A}$ fulfils $\mathbf{G}\mathfrak{q} = \mathbf{G}\mathfrak{f} = \boldsymbol{\mu}_\text{A}$ (orange dot on the right). For each possible deviation from the hypothesis, there is a \totem{plex} (horizontal parallel lines) with associated $I$-projection of $\mathfrak p_\text{H}$ (orange and gray dots). The corresponding statistic $t = 2N\infdiv{\mathfrak q}{\mathfrak{p}_\text{H}}$ with $\mathfrak q=\mathfrak{p}_\text{A},\mathfrak{p}_{\text{A}'}$ is $\chi^2$-distributed (on the left) with the degrees of freedom $k = 1$ being the number of hypothesis conditions. 
		}
	\label{fig:figure1} 
\end{figure}

In the Supplementary Text - Background, we generalize Jaynes' concentration theorem~\cite{Rosenkrantz1989} to the $I$-projection on nested \totem{plices} $\mathcal T_\text{H},\mathcal T_\text{A}\subset \mathcal T_\text{S}$. At large $N$, the sampling probability of datasets with a given $I$-divergence from the hypothesis distribution is \textit{universally} (i.e.\ application-independently) approximated by the $\chi^2$ distribution. In particular, the $p$-value associated to the observed mismatch in the hypothesis conditions can be estimated by the cumulative distribution $1 - F_{\chi^2}(t; k)$ evaluated at the test statistic $t = 2\:N\:\infdiv{\mathfrak p_\text{A}}{\mathfrak{p}_\text{H}}$. Analogously to the hypothesis distribution, $\mathfrak p_\text{A}$ denotes the $I$-projection of $\mathfrak p_\text{H}$ on the alternative \totem{plex} (Fig.~\ref{fig:figure1}). For non-vanishing $p\text{-values}$ in experimental studies, it is $\mathfrak p_\text{A}=\mathfrak{f}$. The degrees of freedom $k$ are fixed by the number of hypothesis conditions, i.e.\ the rank of $\mathbf G$ minus the number of structural conditions.  

Technically, the sampling probability of a \totem{plex} $\mathcal T_{\text A'}\subset\mathcal T_\text{S}$ (i.e.\ the cumulative probability to sample any distribution $\mathfrak p\in\mathcal T_{\text A'}$) that is defined by a certain violation of the hypothesis condition is controlled by $\mathfrak p_{\text A'}=\argmin_{\mathfrak p\in\mathcal T_{\text A'}}\infdiv{\mathfrak p}{\mathfrak p_\text{H}}$. Hence, all the \totem{plices} within structural $\mathcal T_\text{S}$ that are defined by violations with the same value of the controlling $I$-divergence $\infdiv{\mathfrak p_{\text{A}'}}{\mathfrak p_\text{H}}=\infdiv{\mathfrak p_{\text{A}}}{\mathfrak p_\text{H}}$ would share the same sampling probability. This realization gives asymptotically rise to the \totem $\chi^2$ distribution.
To this end, the density $\chi^2_k(t)$ reflects the cumulative probability for all \totem{plices} defined by a violation of the hypothesis condition that is as (im)probable as the observed violation corresponding to $t$.
By its construction thus, our $\chi^2$ distribution automatically provides an ideal candidate to compute the actual $p\text{-value}$. 
 
In total, we arrive at the penultimate pivotal test statistic $t = 2N\infdiv{\mathfrak{p}_\text{A}}{\mathfrak{p}_\text{H}}$, whose value depends on data and theory, but whose distribution is, given the number of hypothesis conditions, application-independent. We refer to the associated statistical hypothesis test as the Information test ($I$-test).

\section*{The $I$-test in the problem of metabolic scaling}

Details about the application of the $I$-test in allometry are given in Supplementary Text - Application.
Kleiber's law~\cite{kleiber_body_1932} relates the body mass $X$ of organisms to their \textsc{bmr} $Y$. Across mammalian species this relationship is often described by a pure allometric law of the form
\begin{equation}
    \label{eqn::kleibersLaw}
    \langle Y\rangle  = a \:\langle X^{3/4}\rangle, 
\end{equation}
where $a >0$ corresponds to a possibly taxon-specific proportionality constant. Expectations allude to the effective character of the law: individual species do not need to obey this relationship given $a$. In fact, there might be many more unaccounted effects which determine \textit{intra}specific traits, as we do not generally know (all) the specifics of data generation. Only after averaging-out intraspecific deviations, we may anticipate the power law to hold between the expectations of observable traits in the \textit{inter}specific population.

A theoretical justification for the scaling exponent $3/4$ is provided by the West, Brown and Enquist model (WBE) linking metabolic scaling to a fractal nutrient-supply network \cite{west_general_1997} such that Kleiber's law ``stands out as genuinely good'' power law \cite{stumpf_critical_2012}.  However, Kleiber's law and its mechanistic explanation by the WBE model has been questioned in the recent years \cite{da_silva_allometric_2006,white_metabolic_2022}. To demonstrate the applicability of \totem and the $I$-test, we set out to test Kleiber's law using \textsc{bmr}-body-mass data for 549 mammalian species \cite{genoud_comparative_2018}. Remaining data-centric, we do not manipulate or further preprocess the data.

Kleiber's law in Equation~\ref{eqn::kleibersLaw} does not \textit{a priori} specify the value of the proportionality constant. Trivially, we could choose $a = \langle Y \rangle_\mathfrak{f} / \langle X^{3/4} \rangle_\mathfrak{f}$ (the ratio of sample averages for $Y$ and $X^{3/4}$), so that the effective law would be fulfilled by any dataset resulting in a no-test scenario. Selecting $a$ to minimize some measure of expected deviation from the power law, such as the mean squared error, would unnecessarily lead to additional bias w.r.t.\ the data. 

To circumvent no-test and biased scenarios without postulating a data generation process parameterized by $a$, we consider the ratio $A=Y/X^{3/4}$ as a random variable whose observations are given by the intraspecific relationships $a_i=y_i/x_i^{3/4}$ for species $i=1,\ldots N$. As the intraspecific proportionality factor varies in an unforeseen way, the expectation of $A$ is not prespecifiable. Therefore, we can at most hypothesize that the ratio is randomly distributed with respect to the explanatory variable $X^{3/4}$. Mathematically, this hypothesis corresponds to a vanishing covariance between $Y / X^{3/4}$ and $X^{3/4}$
\begin{equation}
    \label{eqn::VanishingCovariance}
    \left\langle \left(\frac{Y}{X^{3/4}} - \left\langle \frac{Y}{X^{3/4}}\right\rangle\right)\left(X^{3/4} - \langle X^{3/4} \rangle\right)\right\rangle = 0~,
\end{equation}
which is the minimal condition to test Kleiber's law empirically. It is re-assuring that the hypothesis condition~\ref{eqn::VanishingCovariance} implies the effective form of Kleiber's law,
\begin{equation}
    \label{eqn::allometricLaw_condition}
    \langle Y \rangle = \left\langle \frac{Y}{X^{3/4}}\right\rangle \langle X^{3/4} \rangle ~,
\end{equation}
without any unknowns, exclusively expressed in terms of expectations of observables. The converse does not hold: Kleiber's law for generic value of $a$ in Equation~\ref{eqn::kleibersLaw} does not imply a vanishing covariance between $A$ and $X^{3/4}$. 
Testing Kleiber's law thus reduces to evaluating the probability of observed violation $\langle Y \rangle_\mathfrak{f} - \left\langle A\right\rangle_\mathfrak{f}\: \langle X^{3/4}\rangle_\mathfrak{f}$ in the sample under a population that \textit{exactly} obeys the hypothesis condition in Equation~\ref{eqn::allometricLaw_condition}.

Committing to some fundamental data generation process, the ``old statistics'' lacks a precise explanation for what it means to test an effective law. However, inspired by the effective form in Equation~\ref{eqn::kleibersLaw} of Kleiber's law, we can regress $Y$ against $X^{3/4}$. The fitted value for the intercept corresponds to the observed mismatch $\langle Y \rangle_\mathfrak{f} - \hat{a}\:\langle X^{3/4} \rangle_\mathfrak{f}$ given the fitted value $\hat{a}$ of the proportionality constant. Under the null hypothesis, the intercept must be zero. The ordinary least square estimate of the intercept approximately equals $-358$, which is significantly different from zero ($p = 3.07\times 10^{-5}$; $t$-test). However, a test for normality of the residuals reveals a significant deviation thereof ($p < 2.2\times 10^{-16}$; Shapiro-Wilk normality test). Consequently,  the low $p$-value of the $t$-test may be due to the incompatibility of Kleiber's law with the data and/or due to the unmet assumption of normality required by the $t$-test. Other fitting strategies uncover different estimates for the intercept and slope. However, they cannot mitigate the deviation from normality (Fig. \ref{fig:figureS1}). In summary, the ``old statistics'' provides uninterpretable test results for Kleiber's law applied on the same data. 
 
The $I$-test does not suffer from the problems arising in the ``old statistics''. From Equation~\ref{eqn::allometricLaw_condition}, we  directly write the functions of observables (represented as row vectors in the coefficient matrix $\mathbf G$)
\begin{equation}
    g_\text{N}(X,Y) = 1\quad,\quad
    g_X(X,Y) = X^{3/4}
    \quad\text{and}\quad
    g_0(X,Y) =  Y - \mu_X \frac{Y}{X^{3/4}}~.
\end{equation}
In the ideal population that precisely obeys Kleiber's law, the expectations of these functions are given by $\boldsymbol \mu_\text{H} = \left(1,\mu_X, 0\right)^T$, specifying the linear system in Equation~\ref{eqn::HypothesisLinearSystem} up to the value of the fractional moment $\mu_X$.
Clearly, hypothesis \totem{plex} $\mathcal T_\text{H}$ and hypothesis distribution $\mathfrak p_\text{H}$ depend on the value of $\mu_X>0$. 

In \totem, we have the flexibility to account for (un)certainties about expectations. This enables us to adjust the statistical power of the $I$-test to the study design. Accordingly, expectations in the hypothesis conditions which are known beforehand must be prespecified in order to claim the increased statistical power anticipated with this extra piece of information. 
To prevent ``$p$-hacking'' in this spirit, we should have prespecified $\langle X^{3/4}\rangle$, if we knew its value before data collection. In the current application, this would mean that we controlled the body masses before we measured the corresponding \textsc{bmr}, so that we could estimate the fractional moment of mass from  the sample, $\mu_X = \langle X^{3/4} \rangle_\mathfrak{f}$.

After prespecification of $\mu_X$, the linear system is fully determined. Normalization $g_\text{N}$ and the condition on the fractional moment $g_X$ define the structural \totem{plex} $\mathcal T_\text{S}$.  $I$-projecting $\mathfrak f$ on $\mathcal T_\text{H}$ defines the hypothesis distribution $\mathfrak{p}_\text{H}$. The computation of a $p$-value requires the $I$-projection of the hypothesis distribution $\mathfrak{p}_\text{H}$ onto the alternative \totem{plex} $\mathcal T_\text{A}$ defined by $\mathbf{G}\:\mathfrak{f}$. In absence of zero probabilities, this recovers back the data, $\mathfrak p_\text{A}=\mathfrak f$. By construction, the only difference between $\mathcal{T}_\text{H}$ and $\mathcal{T}_\text{A}$ occurs in the observed violation of hypothesis condition, so that the $p$-value is given by $1 - F_{\chi^2}(2\:N \infdiv{\mathfrak f}{\mathfrak{p}_\text{H}}; 1)$. In this setting, the $I$-test asserts that the observed bias of approximately $ 328$ is significantly different from zero ($p = 0.00187$; $I$-test).  

Actually, the authors who compiled the dataset did not control for the body mass. As a consequence, the fractional moment $\mu_X$ should not be prespecified. It remains unconditioned and must be integrated over its possible manifestations, so that the structural \totem{plex} coincides with the simplex over $\mathcal M$. Numerically, we deal with the corresponding optimization problem under non-linear (in probabilities $p(x,y)$) conditions by parametrizing it via families of $I$-projections. In this scenario, the $p$-value rises to $p = 0.01115$, as a consequence of not knowing the fractional moment $\mu_X$ beforehand, which generically decreases statistical power. However, even in this observational setting, Kleiber's law is rejected by our empirical test at the customary significance level of 5\%. 

Thus, the $I$-test leads to the demise of the effective form of Kleiber's law. Rubner's surface law \cite{rubner_uber_1883}, a pure allometric power law with a scaling exponent of $2/3$, must be rejected due to vanishing $p$-value, as well. Failing in their effective form, the two power laws can be automatically rejected as fundamental laws, as well. In total, the $I$-test reveals that we still lack a widely-accepted mechanistic understanding to explain the relationship between \textsc{bmr} and body mass.

In summary, we demonstrate an application of the $I$-test by empirically testing the allometric power law in different study designs. The \totem framework circumvents the epistemological problem to interpret the results of statistical hypothesis testing which is associated with unwarranted assumptions about the data generation process in the ``old statistics''.

\subsection*{Discussion}
We propose the \totem framework and in particular the $I$-test to allow for statistical testing of scientific theories under the sole assumption of quasi-independent sampling at large sample size $N$. The large-$N$ expansion reveals the minimization of the $I$-divergence as the central element of \totem. Mathematically, it is conceivable to minimize other divergence measures. However, it still remains unclear what limiting processes other divergences might describe. In contrast, there exist many sampling processes whose large-$N$-type limits are governed by the $I$-divergence (Supplementary Text - Background).

Methodologically, the principle of minimal $I$-divergence formally introduces the $\chi^2$ statistic in the pivotal $I$-test. The \totem $\chi^2$ distribution descends from the conditional sampling probability of manifestation distributions, based on general arguments about quasi-independent sampling. Although the defining statistic of the $I$-test in Table~\ref{tab:notation} is proportional to the $I$-divergence, it becomes a complicated, non-linear function of the hypothesized values in attribute space. Working in probability space, we are thus able to streamline the estimation of a totally empirical \text{$p$-value}. The efficient computation of the $I$-projection allows to systematically test families of scientific theories at large scale.

Starting from the postulates of probability theory, Jaynes recognized the universal interpretation of $\chi^2$ statistic, as well. However, his treatment does not relate the universality to the information-theoretic literature of Csiszar et al., as we have done here to justify the validity of the large-$N$ expansion.  
Consequently, the insightful claims of~\cite{jaynes_probability_2003} on the so-called $\psi$-criterion have remained unappreciated so far, not being systematically exploited for hypothesis testing of scientific theories.
Only by expanding the conditional sampling probability around the $I$-projection of the hypothesis, the \totem $\chi^2$ distribution universally arises.

In fact, the conventional view of Pearson's $\chi^2$ test results from an approximation to a special case of the $I$-test. Given some hypothesis distribution, we apply the $I$-test on the alternative \totem{plex} which fully adheres to the data, i.e.\ it includes only $\mathfrak f$, when the structural \totem{plex} coincides with the simplex. Approximating the pivotal statistic in Table~\ref{tab:notation} (for $\mathfrak p_\text{A}=\mathfrak f$) when the data lies close to the hypothesis distribution, we obtain the familiar form of Pearson's $\chi^2$ statistic. Besides the validity of this approximation \textit{per se}, this application of the $I$-test probes all  effects that were observed in the data against a hypothesis about each manifestation probability. As such, it is not appropriate for testing an empirical law which remains agnostic to  effects beyond the expected relationship alongside specifics of the study design, at most. 

Testing Kleiber's law describing the relationship between \textsc{bmr} and body mass demonstrates the usefulness of \totem and the $I$-test. As we outline in Supplementary Text - Application, the $I$-test can be formulated for further scenarios, like testing the equality of group means or the equality of group rates in settings where the group sizes are either fixed (in experimental studies) or \textit{a priori} unknown (in observational studies).

Our presentation of the $I$-test adheres very closely to the established logic of hypothesis testing. In particular, we still argue along the lines of the probability of the data given a hypothesis. Such reasoning must not be associated with evidence \emph{for} a scientific theory in form of posterior probabilities of hypotheses given the data. Lacking some clear mechanistic understanding of the underlying process, we can conceive uncountably many hypotheses relating some attributes. Since the probability of a single hypothesis differentially vanishes, 
evidence in form of a posterior probability of a hypothesis given the data is ill-defined. Naively, we could compare hypotheses via the $I$-divergence of their hypothesis distributions from the data. However, it is \textit{a priori} clear that the ``best'' hypothesis with lowest $I$-divergence from $\mathfrak{f}$ is trivially $\mathfrak{f}$ itself!

\totem and the $I$-test do not pose as an oracle that will automatically provide the ``correct'' theory from data. Despite the optimism in the data-scientific community, we strongly believe that observations alone --\,lacking deeper domain knowledge and intuition\,-- are far from sufficient to discover scientific theories. Nevertheless, \totem and the $I$-test provide a robust statistical framework to efficiently test a scientific ansatz on the given data avoiding unwarranted distributional assumptions on random variables. In that way, the researcher remains only concerned with the core hypothesis deducible from the theory without worrying about (theory-unrelated) constraints imposed by existing statistical models. Enabling statistical testing of versatile theories in a totally empirical manner, we thus provide a totally empirical basis of science.
 


\begin{table} 
	\centering
	\caption{\textbf{Concepts and Definitions.}}
	\label{tab:notation} 
	
	\begin{tabular}{ll} 
		\\
		\hline
		Concept & Symbol/Definition\\
		\hline \hline
		physical scale & sample size $N$\\
		quasi-independent sampling & $N\gg1$ (statistical regime) \\
		\hline
		attributes & $X,Y$\\
		joint manifestation & $(x,y)$\\
		set of observed joint manifestations & $\mathcal{M}$\\
		probability of joint manifestation & $p(x,y)$ \\
		probability simplex & $\mathcal{P} = \left\{\mathfrak{p} \in \mathbb{R}_{\ge 0}^{\vert \mathcal{M}\vert} ~\vert~\sum_{(x,y) \in \mathcal{M}} p(x,y) = 1\right\}$\\
		\hline
		function of attributes & $g_a(X,Y)$\\
		expectation of function given $\mathfrak{p}$ & $\langle g_a(X,Y) \rangle_\mathfrak{p} = \sum_{(x,y)\in\mathcal M}g_a(x,y)\,p(x,y)$\\
		hypothesized value & $\mu_a\in\mathbb R$ \\
		hypothesis \totem{plex}& $\mathcal{T}_\text{H} = \left\{ \mathfrak{p} \in \mathcal{P} ~\vert~ \vec g_a\cdot\mathfrak{p} = \mu_a\,,~a=1,\ldots D\right\}$ \\
		alternative \totem{plex} & $\mathcal{T}_\text{A}= \left\{ \mathfrak{p} \in \mathcal{P} ~\vert~ \vec g_a\cdot\left(\mathfrak{p}-\mathfrak{f}\right)=0\,,~a=1,\ldots D\right\}$\\
		structural \totem{plex} & $\mathcal{T}_\text{S}\supset \mathcal T_\text{H},\mathcal T_\text{A}$ \\
		\hline
		$I$-divergence & $\infdiv{\mathfrak{p}}{\mathfrak{u}}=\sum_{(x,y) \in \mathcal{M}} p(x,y)~\log\frac{p(x,y)}{u(x,y)}$ \\
		hypothesis distribution & $\mathfrak{p}_\text{H} = \argmin_{\mathfrak p\in\mathcal{T}_\text{H}}\infdiv{\mathfrak p}{\mathfrak f}$ \\
		alternative distribution & $\mathfrak{p}_\text{A}= \argmin_{\mathfrak p\in\mathcal{T}_\text{A}}\infdiv{\mathfrak p}{\mathfrak{p}_\text{H}}$ \\
		$\chi^2$ statistic of $I$-test & $t = 2~N~\infdiv{\mathfrak{p}_\text{A}}{\mathfrak{p}_\text{H}}$ \\
		degrees of freedom of $\chi^2$ & $\text{\# hypothesis conditions} = D - \text{\# structural conditions}$\\
		\hline
	\end{tabular}
\end{table}


\clearpage 

%
\bibliography{totem} 
\bibliographystyle{sciencemag}

%
%
%
%
%
%


\section*{Acknowledgments}
We thank Hans-Helge M\"uller for introducing Barnard's test to us. We thank Till Adhikary and Petra Fischer for proof-reading the manuscript. 
\paragraph*{Funding:}
O.L. and H.-R.C. were solely funded by the Philipps-University Marburg.
\paragraph*{Author contributions:}
Conceptualization: O.L. and H.-R.C.; Formal analysis: O.L. and H.-R.C.; Software: O.L. and H.-R.C.; Writing – original draft: O.L.; Writing – review and editing: O.L. and H.-R.C.
\paragraph*{Competing interests:}
There are no competing interests to declare.
\paragraph*{Data and materials availability:}
The \textsc{totem} R-package is available at \url{https://github.com/horyunchung/totem}. The package includes the processed \textsc{bmr} and body mass data. The application of the $I$-test for testing Rubner's and Kleiber's law is exemplified in the package vignette.

\subsection*{Supplementary materials}
Materials and Methods\\
Supplementary Text - Background\\
Supplementary Text - Application \\
Fig. S1\\
References \textit{(31-\arabic{enumiv})}\\ 


\newpage


\renewcommand{\thefigure}{S\arabic{figure}}
\renewcommand{\thetable}{S\arabic{table}}
\renewcommand{\theequation}{S\arabic{equation}}
\renewcommand{\thepage}{S\arabic{page}}
\setcounter{figure}{0}
\setcounter{table}{0}
\setcounter{equation}{0}
\setcounter{page}{1} 


\begin{center}
\section*{Supplementary Materials for\\ \scititle}

Orestis Loukas,
Ho-Ryun Chung$^\ast$ \\
\small$^\ast$Corresponding author. Email: ho.chung@staff.uni-marburg.de\\
\end{center}

\subsubsection*{This PDF file includes:}
Materials and Methods\\
Supplementary Text - Background\\
Supplementary Text - Application\\
Figure S1\\


\newpage


\subsection*{Materials and Methods}
\subsubsection*{Data Import}
We downloaded the basal metabolic rate (\textsc{BMR}) database from Table S2 and the mammalian phylogeny from File S1 \cite{genoud_comparative_2018}. Since File S1 was a PDF file, we copied its contents into a text file (\texttt{mammalian.nexus}). All analysis was performed in \textsc{R} (v4.3.1; \cite{R_language_2023}). We imported the \textsc{BMR} database into \textsc{R} using the function \texttt{read\_xlsx} from the \textsc{readxl} package (v1.4.3; \cite{wickham_readxl_2023}).
\begin{verbatim}
data <- data.frame(
    read_xlsx(
        path = 'brv12350-sup-0003-tables2.xlsx', 
        skip = 2, 
        na = "NA"
    ), 
    check.names = FALSE
)
\end{verbatim}
and the nexus tree using the function \texttt{read.nexus} from the \textsc{ape} package (v5.8; \cite{paradis_ape_2019})
\begin{verbatim}
phy <- read.nexus('mammalian.nexus')
\end{verbatim}

From the BMR database we selected only rows, where the ``SELECT'' column contained the value 1 and retained only the columns ``Species Nexus'', ``BMR'', ``body mass''.
\begin{verbatim}
data <- subset(
    x = data, 
    subset = SELECT == 1, 
    select = c("Species Nexus", "BMR", "body mass")
)
\end{verbatim}
We renamed the columns
\begin{verbatim}
colnames(data) <- c("Species", "BMR", "Mass")
\end{verbatim}
and reformatted the species names to match the species names in the nexus file
\begin{verbatim}
data$Species <- gsub("\\s", "\\_", data$Species) 
\end{verbatim}
For the phylogenetic generalized linear model fitting \cite{orme_caper_2023} we construct a \texttt{comperative.data} object
\begin{verbatim}
comperativeData <- comparative.data(
  data = data, 
  phy = phy, 
  names.col = "Species"
)
\end{verbatim}

\subsubsection*{Data Analysis}
\paragraph*{Ordinary Least Square Regression} We used the \texttt{lm} function to fit Rubner's law \cite{rubner_uber_1883} or Kleiber's law \cite{kleiber_body_1932}, i.e. allometric laws $Y = a^{(\text{b})}\:X^\text{b}$ with the exponent $\text{b} = 2/3$ and $\text{b} = 3/4$, respectively.
\begin{verbatim}
results <- list()
results$lm <- list(
    rubner = lm(BMR ~ I(Mass^(2/3)), data = data),
    kleiber = lm(BMR ~ I(Mass^(3/4)), data = data)
)
\end{verbatim}
\paragraph*{Major Axis, Standard Major Axis, and Ranged Major Axis Regression} We used the \texttt{lmodel2} function from the package \textsc{lmodel2} (v1.7-3; \cite{legendre_lmodel2_2018})
\begin{verbatim}
results$lmodel2 <- list(
    rubner = lmodel2(
        BMR ~ I(Mass^(2/3)), 
        data = data, nperm = 1000, 
        range.y = "relative", 
        range.x = "relative"
    ),
    kleiber = lmodel2(
        BMR ~ I(Mass^(3/4)), 
        data = data, nperm = 1000, 
        range.y = "relative", 
        range.x = "relative"
    )
)
\end{verbatim}
\paragraph*{Phylogenetic Generalized Linear Model}
We used the function \texttt{pgls} from the package \textsc{caper} (v1.0.3; \cite{orme_caper_2023})
\begin{verbatim}
results$pgls <- list(
    rubner = pgls(
        BMR ~ I(Mass^(2/3)), 
        data = comperativeData, 
        lambda = "ML"
    ),
    kleiber = pgls(
        BMR ~ I(Mass^(3/4)), 
        data = comperativeData, 
        lambda = "ML"
    )
)
\end{verbatim}
\paragraph*{$I$-test} We used the function \texttt{i.test} from the package \textsc{totem} (v1.0; this study)
\begin{verbatim}
results$totem <- list(
  rubner = i.test(BMR ~ I(Mass^(2/3)), data = data),
  kleiber = i.test(BMR ~ I(Mass^(3/4)), data = data)
)
\end{verbatim}



\subsection*{Supplementary Text - Background} 

\paragraph{The algebra of empiricism}
In this work, we embark upon the most empirical description that is quantitatively possible. No prior distributional assumptions are made for random variables $X^\alpha$  representing observable attributes $\alpha=1,\ldots,L$ in the analysis. In other words, we do not assume a concrete statistical model for the generation of data.
The values of random variables can come from categorical or metric domains $\mathcal X^\alpha$.  
An empirical description that does not commit to any model is built on joint manifestations $\underaccent{\bar}{x}\equiv(x^1, \ldots, x^\alpha, \ldots, x^L)$ specifying the value  $x^\alpha\in \mathcal X^\alpha$ for each attribute. We denote by  $\mathcal M$ the set of possible manifestations.

In principle, the cardinality of $\mathcal M$ can exponentially grow due to combinatorics over categorical attributes or be infinite due to metric attributes.  However, the size $N_\text{real}$ of realistic datasets imposes an upper bound $\vert\mathcal M\vert\leq N_\text{real}$ that is observed far below the theoretically possible limit. 
Hence, the set of relevant manifestations will be a subset of the Cartesian product of attribute domains, $\mathcal M\subseteq \mathcal X^1 \times\cdots\times \mathcal X^L$. Since only a finite number of values are observable in a finite dataset $N_\text{real}<\infty$, we interchangeably use the product notation also for the domain of a metric variable.

In contrast to the ``old statistics'' that analyzes the joint distribution of random variables based on assumed special traits (normality, skewness etc.), \totem focuses on the probability $p_x\equiv p(\underaccent{\bar}{x})$ with which each  distinct $\underaccent{\bar}{x} \in\mathcal M$ occurs in a given setting.
The probabilities of all manifestations are collected in a distribution $\mathfrak p:\,\mathcal M\rightarrow[0,1]^{\vert \mathcal M\vert}$.
Generalizing the idea of the simplex, we represent distributions $\mathfrak p$ as column vectors with entries on the \totem{plex}
\begin{equation}
\label{eq:TOTEMplex:concept}
\mathcal T = \left\{\left.\mathfrak p \in \mathbb R_{\geq0}^{\vert\mathcal M\vert}  ~\right\vert~ \text{conditions on $p_x$} \right\}~.
\end{equation}
Manifestations themselves --\,arbitrarily enumerated, since their ordering bares no physical meaning\,-- can be thought of as spanning the standard orthonormal basis $\{\mathbf e_x\}_{\underaccent{\bar}{x}\in\mathcal M}$ in $\mathbb R^{\vert\mathcal M\vert}$.  

A well-defined dataset $\boldsymbol D$ is provided as a two-dimensional table with the $i$-th data point represented by a complete row that specifies the observed value $D_i^{~\alpha}\in\mathcal X^\alpha$ of each attribute $\alpha$ in the analysis. Clearly, there might be many redundant rows in $\boldsymbol D$ whose abundance can be neatly summarized via a histogram, after mapping the rows of $\boldsymbol D$ to manifestations from $\mathcal M$. 
Since extrapolations are meaningless whenever lacking an underlying model of data generation, we are only interested in observed manifestations for which a non-vanishing histogram frequency\footnote{For each attribute $\alpha$, the indicator map $\delta: \mathcal X^\alpha\times\mathcal X^\alpha\rightarrow \{0,1\}$ is defined over the corresponding domain.}
\begin{equation}
\label{eq:PositiveHistogramFrequency}
0~ <~ Nf_x = \sum_{i=1}^N \prod_\alpha\delta(D_i^{~\alpha}, x^\alpha)
\end{equation}
has been recorded from $N$ data points. All non-vanishing histogram relative  frequencies will be denoted by the empirical distribution $\mathfrak f$ with $f_x = \mathbf e_x\cdot \mathfrak f$ in the standard orthonormal basis.

For the purposes of empirical testing, the \totem{plex} in Equation~\ref{eq:TOTEMplex:concept} needs to be formally defined over observed manifestations which remain stochastically possible: 
\begin{equation}
\label{eq:admissibleEntities}
\forall\: \underaccent{\bar}{x}\in \mathcal M\,:\quad\exists\, \mathfrak p\in\mathcal T \quad{\text{s.t.\ }}\quad p_x>0 ~.
\end{equation}
In contrast, we exclude from the onset any theoretically possible manifestation whose probability always vanishes in the \totem:
\begin{equation}
\label{eq:ZeroConditions}
\not\exists\: \underaccent{\bar}{x}\in\mathcal M \quad{\text{s.t.\ }}\quad p_x=0 ~~\forall\, \mathfrak p \in\mathcal T~.
\end{equation}

In this paper, we express investigated relationships and conjectured hypotheses via $D$ linear conditions on $\mathfrak p\in\mathcal T$ with concrete physical meaning,
\begin{equation}
\label{eq:LinearSystem}
\mathbf G \, \mathfrak p = \boldsymbol\mu
~,
\end{equation}
controlled by some real-valued vector $\boldsymbol\mu\in\mathbb R^D$ which we specify application-wise. Each condition row-vector $\vec g_a\in\mathbb R^{1\times\vert\mathcal M\vert}$ of the coefficient matrix $\mathbf G$ prescribes how to calculate the expectation of some quantity (a function of random variables). The \totem framework assigns to expectations fixed values using the standard scalar product in $\mathbb R^{\vert\mathcal M\vert}$:
\begin{equation}
\label{eq:Expectation}
\vec g_a \cdot \mathfrak p = \mu_a \quad\text{for}\quad a=1,\ldots,D
~.
\end{equation}
$\mu_a\in\mathbb R$ can reflect the value of any observed or hypothesized moment, marginal or expectation of a more general function of random variables. For example, the expected body-mass index in terms of random variables  height $H$ and weight $W$ related via $g(H,W)=H^{-2}W$ would be calculated according to 
\begin{equation}
\nonumber
\left\langle H^{-2}\,W\right\rangle_{\mathfrak p} := \sum_{h,w} p(h,w) \frac{w}{h^2}~,
\nonumber
\end{equation}
using tuples $(h,w)\in\mathcal M$ to parameterize the entries of our probability vectors.  
 
Evidently, linear-algebraic expressions of this form make sense, because there are only countably many observed manifestations of attributes in a finite dataset. In particular, normalization on the simplex 
\begin{equation}
\langle 1 \rangle_{\mathfrak p} := \sum_{\underaccent{\bar}{x}\in\mathcal M} p_x = 1
\end{equation}
will be always implied on $\mathcal T$, as well.  In that way, the \totem{plex} describes a subregion of the conventional simplex over manifestations $\mathcal M$. In the following, the coefficient matrix $\mathbf G\in\mathbb R^{D\times\vert\mathcal M\vert}$ is assumed to be of full row-rank $D=\rank\mathbf G$, after the necessary algebra has eliminated any redundancies.

Given the same column space over manifestations, two coefficient matrices related by full-rank transformation matrix $\mathbf S\in\mathbb R^{D\times D}$ via 
\begin{equation}
\label{eq:TOTEM:Reparametrization}
\mathbf G \quad\rightarrow\quad \mathbf S\, \mathbf G
\end{equation}
construct the same \totem{plex}.
If linear conditions in Equation~\ref{eq:LinearSystem} admit at least one solution, (e.g.\ the empirical distribution itself whenever $\boldsymbol\mu = \mathbf G\:\mathfrak f$), there exist infinitely many distributions on the \totem{plex}
generated by convexity through
\begin{equation}
\label{eq:TOTEMplex:Convexity}
\lambda{\mathfrak p} + \left(1-\lambda\right) {\mathfrak p'}\in\mathcal T
\quad\text{for}\quad
\mathfrak p, \mathfrak p'\in\mathcal T
\quad\text{and}\quad\lambda\in[0,1]~,
\end{equation}
where $(\mathfrak p - \mathfrak p')\in\ker\mathbf G$, 
as long as $D < \vert\mathcal M\vert$, so that $\vert\ker\mathbf G\vert=\vert\mathcal M\vert-D>0$.
In that way, 
$\mathcal T$ defines an equivalence class on the ambient simplex, as its member  distributions share the same expectations dictated by control vector $\boldsymbol\mu$. 

Furthermore, two \totem{plices} whose coefficient matrices  $\mathbf G$ and $\mathbf G'$ are related  for $D<D'$  through
    \begin{equation}
        \label{eq:NestedTOTEMs:CoefficientMatrix}
        \mathbf G = \mathbf S\: \mathbf G'
    \end{equation}
    given $\mathbf S\in\mathbb R^{D\times D'}$
    with $\rank\mathbf S=D$ (otherwise $\mathbf G$ would be row-rank deficit)
    are called nested, as long as controlling conditions
    \begin{equation}
    \label{eq:NestedTOTEMs:moments}
    \boldsymbol \mu = \mathbf S\, \boldsymbol \mu'
    \end{equation}
    transform covariantly under Equation~\ref{eq:NestedTOTEMs:CoefficientMatrix}.
The induced \totem{plices} are subsets, $\mathcal T'\subset\mathcal T$, as can be readily seen from linearity of expectations in matrix-vector notation: 
\begin{align*}
\mathfrak p\in\mathcal T'\,:\quad
\mathbf G\, \mathfrak p = \mathbf S\: \mathbf G' \mathfrak p = \mathbf S\: \boldsymbol \mu' = \boldsymbol \mu 
\quad\Rightarrow\quad
\mathfrak p\in\mathcal T
~.
\end{align*}

\paragraph{Conditional sampling at large sample size} 
Besides theory-related $\vert\mathcal M\vert$ and $D=\rank\mathbf G$, the only physical scale in a \totem analysis is the sample size $N$, as little should be assumed about underlying processes that contributed to data generation. At sufficiently large $N$, many quasi-independent sampling processes cannot be effectively distinguished from multinomial sampling~\cite{csiszar_sanov_1984}. Below, we provide such large-$N$-type limits for familiar sampling processes that approximately reduce to multinomial sampling, in spite of not formally consisting of independent events. 

In the so-called \textit{universality class} of quasi-independent sampling, we remain sufficiently generic in the statistical regime, even after committing to one concrete sampling process. We choose the  multinomial process which specifies the minimal amount of information that effectively characterizes the universality class: intrinsic probabilities $u_x$ (our bias) and a count vector $\mathbf n$ (the sampled histogram). 
Eventually, we specify the reference distribution $\mathfrak u$ relevant for \totem hypothesis testing. 
For now, we note that the recorded data would have been impossible, if any entry in $\mathfrak u$ were zero. We emphasize that the reference distribution could (and most likely would) fail to satisfy linear system Equation~\ref{eq:LinearSystem}, i.e.\ generically $\mathfrak u\not\in\mathcal T$.

Given a reference distribution, the object to investigate in the universality class of quasi-independent sampling is the conditional probability 
\begin{equation}
\label{eq:MutlinomialDistro}
\text{Pr}(N\mathfrak p\vert \mathcal T;\mathfrak u) = Z^{-1} \text{mult}\left(\mathbf n; \mathfrak u\right)
~,
\end{equation}
where sampled counts 
\begin{equation}
\label{eq:from_Counts_to_Distros}
\mathbf n = N\mathfrak p \in \mathbb N^{\vert\mathcal M\vert}~,
\end{equation}
are subject to the conditions of Equation~\ref{eq:LinearSystem}. Extending the Diophantine system over rational vectors $\mathbf n/N$ onto a linear system of coupled equations over the non-negative reals we can approximate its solution space, the lattice polytope, by the \totem{plex} in Equation~\ref{eq:TOTEMplex:concept}.
The normalizing factor in Equation~\ref{eq:MutlinomialDistro} is the partition sum
\begin{equation}
\label{eq:MultinomialSampling:PartitionFunction}
Z = \sum_{\mathfrak p\in\mathcal T} \text{mult}(N\mathfrak p;\mathfrak u)~,
\end{equation}
whose continuous approximation reflects the total probability mass of the \totem{plex}. Henceforth, we refer to a sampled dataset comprising the sampled frequencies $n_x$ of manifestations via the associated distribution $\mathfrak p$ in Equation~\ref{eq:from_Counts_to_Distros}.
 
The large-$N$ expansion of the conditional sampling probability in the universality class follows the traditional paradigm of Boltzmann-Sanov's large deviation theory~\cite{sanov_probability_1958}, although the latter addresses the concrete process of sampling with replacement. Concretely, the probability density to sample the differential region around $\mathfrak p\in\mathcal T$  scales by Stirling's formula at $N\gg\vert{\mathcal M}\vert$  as
\begin{equation}
\label{eq:MutlinomialSampling:LargeN}
\log
\text{Pr}(N\mathfrak p\vert\mathcal T;\mathfrak u) 
=
- N \infdiv{\mathfrak p}{\mathfrak u} 
-\tfrac12 \sum_{\underaccent{\bar}{x}\in\mathcal M} \log p_x 
\, + \, \mathcal O\left(N^{-1}\right) -\log  Z
~.
\end{equation}
Therefore, the large-$N$ expansion reveals that sampling from reference distribution $\mathfrak u$ is governed by the $I$-divergence of distribution $\mathfrak p\in\mathcal T$ from  $\mathfrak u$, 
\begin{equation}
\label{eq:iDivergence}
\infdiv{\mathfrak p}{\mathfrak u}  
=\sum_{\underaccent{\bar}{x}\in\mathcal M} p_x\log \frac{p_x}{u_x} 
~. 
\end{equation}
This is the non-negative (by Gibbs' inequality) measure that emerges at large $N$ to govern the sampling probability in the class of quasi-independent processes. Shannon's entropy is only a special case, discussed around Equation~\ref{eq:Idiv_Entropy:Relation}.

Since $\mathfrak p$-independent terms are not relevant for the variational methods we consider in the following, we can silently absorb them into the partition sum ($Z$ stays finite, as it is defined over observed manifestations only).
For the purposes of hypothesis testing, the actual value of $Z$ is not interesting, as we are concerned with the conditional sampling probability that is effectively governed by the the variational extremum of $\text{Pr}(N\mathfrak p\vert \mathcal T;\mathfrak u)$.

At this point, we review the large-$N$-type expansion of well-known sampling processes that belong to the universality class of quasi-independent sampling. For brevity, we allow the \totem{plex} to coincide with the simplex over observed manifestations, as the effect of conditions Equation~\ref{eq:LinearSystem} modify the expansion of the partition function which is however $\mathfrak p$-independent. Ultimately, the variational problem of conditional minimization of the $I$-divergence (presented in the next paragraph) will enforce sampled counts in the leading order of Equation~\ref{eq:MutlinomialSampling:LargeN} to obey the linear system in expectation and the latest asymptotically. 

In the realm of hypergeometric sampling (without replacement) from a finite population of size $K$, 
Fisher's noncentral distribution describes a process of biased --\,through weights $\boldsymbol w\in \mathbb R_+^{\vert{\mathcal M}\vert}$, but still independent\,-- sampling:
\begin{align*}
\text{Pr}\left(N\mathfrak p; K\mathfrak q, \boldsymbol w\right) = Z^{-1}\prod_{\underaccent{\bar}{x}\in\mathcal M}\binom{K q_x}{N p_x} w_x^{Np_x}
\end{align*}
$\mathfrak q$ denotes the population distribution.
By using Stirling's formula to expand each binomial coefficient (under $Kq_x>Np_x$), 
\begin{equation}
    \label{eq:LargMNexpansion:Hypergeometric}
    \log\text{Pr}\left(N\mathfrak p; K\mathfrak q, \boldsymbol w\right) = 
    - N \infdiv{\mathfrak p}{\mathfrak u} 
    -\tfrac12 \sum_{\underaccent{\bar}{x}\in\mathcal M} \log p_x 
     + \mathcal O\left(N^{-1}\right) + \mathcal O\left(N^2/K\right) -\log  Z~,
\end{equation}
it is readily recognizable that the expansion coincides at $K\gg N^2\gg 1$ with the multinomial expansion in Equation~\ref{eq:MutlinomialSampling:LargeN},
with the arising reference  $\mathfrak u$ given by the product (in fact $\mathfrak u$ need not be properly normalized, see~\cite{csiszar_i-divergence_1975})
\begin{equation}
\label{eq:hypergeometric:effective_hypothesis}
u_x \equiv w_x \cdot q_x~.
\end{equation}
Clearly, when all weights are the same, the process reduces to the large $K$-$N$ expansion of the standard, unbiased hypergeometric sampling. 

The Wallenius' noncentral distribution is another example of biased hypergeometric sampling with competing sampling of $N$ individuals from a population of size $K$:
\begin{align*}
\text{Pr}\left(N\mathfrak p; K\mathfrak q, \boldsymbol w\right) = 
\left[\prod_{\underaccent{\bar}{x}\in\mathcal M} \binom{Kq_x}{Np_x} \right]\int_0^1 d t \prod_{\underaccent{\bar}{x}\in\mathcal M} \left(1 - t^{w_x/D}\right)^{Np_x}
\quad{\text{where}}\quad
D = \sum_{\underaccent{\bar}{x}\in\mathcal M} w_x \left(Kq_x-Np_x\right)~.
\end{align*}
Performing the $t$-integral order-by-order in $N^2/K\ll1$,
we  arrive at the  expansion Equation~\ref{eq:LargMNexpansion:Hypergeometric}  with the reference distribution again given by the identification in Equation~\ref{eq:hypergeometric:effective_hypothesis}.

A frequently invoked conjugate prior for multinomial sampling is the Dirichlet distribution with a parameter vector $\boldsymbol\alpha\in\mathbb R_+^{\vert{\mathcal M}\vert}$. Setting 
\begin{equation*}
K = \sum_{\underaccent{\bar}{x}\in\mathcal M}\alpha_x
\end{equation*}
the probability to sample a dataset of size $N$ from the compound Dirichlet-multinomial distribution is given by 
\begin{equation*}
\text{Pr}(N\mathfrak p; \boldsymbol \alpha) = 
\int_{\mathfrak q\in\mathcal P} \text{mult}(N\mathfrak p; \mathfrak q) \, \text{Dir}(\mathfrak q;\boldsymbol\alpha)
=\frac{\Gamma(K)\, \Gamma(N+1)}{\Gamma(K+N)} \prod_{\underaccent{\bar}{x}\in\mathcal M} \frac{\Gamma(Np_x+\alpha_x)}{\Gamma(\alpha_x)\,\Gamma(Np_x+1)}
\end{equation*}
in terms of gamma functions. 
Reparametrizing as $\boldsymbol\alpha = K(\boldsymbol\alpha/K) \equiv K\mathfrak u$, we localize the compound integral around the prior $\mathfrak u$ by taking the $K\gg N^2\gg 1$ limit to obtain the expansion of Equation~\ref{eq:LargMNexpansion:Hypergeometric}.

Next, we consider multiple coupled Poisson processes. Recall that a single Poisson process in a given time interval  describes counts independently reaching the detector at constant mean rate. A simple violation of independence is achieved by a process where counts $m_x\in\mathbb N$ for recorded manifestations are generated based on independent Poisson processes, but one silent Poisson process emitting $m_0\in\mathbb N$ contaminates the recorded signal, so that our detectors provide dependent counts 
\begin{equation*}
n_x = m_x + m_0 \quad\text{for}\quad \underaccent{\bar}{x}\in\mathcal M
~.
\end{equation*}
Detecting a lot of counts, i.e.\ $\mathbf n = N \mathfrak p$ as before, 
it might be reasonable to assume a large-$N$ scaling of the Poisson mean counts $\langle \boldsymbol m\rangle$ against a finite-size mixing effect. In that scenario, the parameters of the participating Poisson distributions would scale as
\begin{equation*}
 \langle m_x\rangle = N u_x \quad\text{and}\quad   \langle m_0 \rangle=\mathcal O(1)~.
\end{equation*}
According to~\cite{Karlis_Mixture_2005}, the sampling distribution for the observed counts is given by the closed form
\begin{equation*}
\text{Pr}(N\mathfrak p; N\mathfrak u, m_0) = e^{-N-\langle m_0\rangle} N^N \left(\prod_{\underaccent{\bar}{x}}\frac{u_x^{Np_x}}{(Np_x)!}\right)
\sum_{\nu=0}^N \left(\prod_{\underaccent{\bar}{x}}\binom{Np_x}{\nu}\right) \nu! \left(\frac{\langle m_0\rangle}{\prod_{\underaccent{\bar}{x}'}Nu_{x'} }\right)^\nu
~.
\end{equation*} 
The mixing-implementing summation over $\nu$ scales as $1+\langle m_0\rangle \prod_{\underaccent{\bar}{x}} p_x/u_x + \ldots$, so that the $\mathfrak p$-dependent part of the large-$N$ expansion agrees with Equation~\ref{eq:MutlinomialSampling:LargeN}
up to $\mathcal O(1)$-corrections from the mixing.
Incidentally, the infinite support of Poisson distribution does not interfere with the joint asymptotic limit for sampled and mean counts.
Even if counts were sampled from a finite ``population'' in Poisson-type processes, 
\begin{equation*}
N p_x \in \{0,\ldots, K q_x\} \quad\text{with}\quad \sum_{\underaccent{\bar}{x}\in\mathcal M} q_x = 1~,
\end{equation*} 
the large-$N$ expansion would still coincide with Equation~\ref{eq:MutlinomialSampling:LargeN} up to finite-support effects being exponentially suppressed by $K\gg N\gg1$.
A similar argumentation shows that the negative binomial distribution (also defined over infinite support) exhibits a large-$N$-type limit, as well. 

For Poisson distributed counts $N\mathfrak p$, one alternatively considers conjugate priors by gamma distributions with shape parameters $K\boldsymbol\alpha$ and scale parameters (inverse rates) $NK^{-1}\boldsymbol\beta$, 
so that the front-end process computes to 
\begin{equation*}
\text{Pr}(N\mathfrak p; \boldsymbol\alpha, \boldsymbol\beta) = \prod_{\underaccent{\bar}{x}\in\mathcal M} \left(\frac{N \beta_x}{K}\right)^{Np_x}  \left(1+\frac{N \beta_x}{K}\right)^{-K\alpha_x-Np_x} \frac{\Gamma\left(K\alpha_x+Np_x\right)}{\Gamma\left(K\alpha_x\right)(Np_x)!}
~.
\end{equation*} 
In this parametrization, we take $\mathcal O(\boldsymbol\alpha)=1$ and $\mathcal O(\boldsymbol\beta)=1$, so that gamma priors localize at 
\begin{equation*}
\langle\lambda_x\rangle = \int_0^\infty \lambda_x\: \text{gamma}(\lambda_x;\alpha_x,\beta_x) = N \alpha_x\beta_x 
\end{equation*}
under a variance
\begin{equation}
\label{label:GammaDistro:Variance}
\langle\lambda_x^2\rangle - \langle\lambda_x\rangle^2= \frac{N^2}{K} \alpha_x\beta_x^2~.
\end{equation}
At $K\gg N^2\gg1$ thus, the gamma priors localize the mean event rates of Poisson processes at $\mathcal O(\langle\boldsymbol\lambda\rangle)=N$, with sub-leading Poisson processes quickly suppressed by Equation~\ref{label:GammaDistro:Variance},
showing  that this limit of gamma-Poisson compound distribution falls again within the universality of Equation~\ref{eq:MutlinomialSampling:LargeN}.
Physically, we consider the limit of  a very large number $\mathcal O(K)$ of driver events necessary for  the main event to take place, each sub-event happening within a  small $\mathcal O(N/K)$ time interval (conversely at large sub-event rate), such that in total it takes on average $\mathcal O(N)$ time for the main event.

All in all, we see that various sampling processes which violate the independence assumption in specific ways cannot be distinguished in their specifics from each other, once the large $N$-type limit applies. In this limit, the $I$-divergence of sampled datasets from the reference distribution universally determines all these processes.

\paragraph{The information projection} 
A theorem of paramount importance to formulate totally empirical statistical testing concerns the existence and uniqueness of a special distribution on the \totem{plex}. This fact  was conclusively proven by Csiszár~\cite{csiszar_i-divergence_1975}.
Given a consistent linear system defining a \totem{plex} $\mathcal T$ alongside some reference distribution $\mathfrak u$ (not necessarily belonging to $\mathcal T$), there always exists a unique distribution $\mathfrak q\in\mathcal T$ with  minimum $I$-divergence from the reference. Mathematically, the information projection ($I$-projection) satisfies
\begin{equation}
\label{eq:iProjection}
\mathfrak q=\argmin_{\mathfrak p\in\mathcal T}\infdiv{\mathfrak p}{\mathfrak u}
\quad{\Leftrightarrow}\quad \infdiv{\mathfrak p}{\mathfrak u} \geq \infdiv{\mathfrak q}{\mathfrak u}\quad\forall\,\mathfrak p\in\mathcal T~.
\end{equation}

To sketch the proof we recycle familiar notions of information geometry, see e.g.~\cite{amari_methods_2000}. The uniqueness of a minimum of $I$-divergence immediately follows in probability space by the convexity of the feasible region $\mathcal T\subset\mathcal P$ of the linear  problem Equation~\ref{eq:LinearSystem} at hand, combined with strict convexity of the $I$-divergence thought as a function $[0,1]^{\vert{\mathcal M}\vert}\rightarrow\mathbb R_0^+$ in its first argument. Hence, the $I$-divergence Equation~\ref{eq:iDivergence} possesses one global minimum in $\mathcal T$, at most. At the same time, a \totem{plex} can be thought of  as a non-empty, convex, bounded and closed --\,hence compact\,-- subset of $[0,1]^{\vert\mathcal M\vert}$ over which any continuous function necessarily attains a minimum by the extreme value theorem. In total, we conclude that the $I$-divergence must attain its global minimum, the $I$-projection, in the \totem{plex}.
In the following, we shall make excessive use of the well-known
``Pythagorean'' lemma for $I$-divergences inside the \totem{plex},
\begin{equation}
\label{eq:Pythagoras}
\mathfrak q=\argmin_{\mathfrak p\in\mathcal T}\infdiv{\mathfrak p}{\mathfrak u} \quad{\Leftrightarrow}\quad
\infdiv{\mathfrak p}{\mathfrak u} = \infdiv{\mathfrak q}{\mathfrak u} + \infdiv{\mathfrak p}{\mathfrak q} 
~~\forall\, \mathfrak p\in \mathcal T~,
\end{equation}
proven in a multitude of textbooks on information theory and geometry.

On the boundaries of the \totem{plex}, there exist ``extreme'' distributions $\mathfrak b$ where the probability $b_x$ of at least one manifestation in Equation~\ref{eq:admissibleEntities} vanishes.
Using some distribution $\mathfrak p\in\mathcal T$ we introduce --\,exploiting the convexity Equation~\ref{eq:TOTEMplex:Convexity}\,-- an one-parameter family of distributions 
\begin{equation}
\mathfrak{l} = {\mathfrak b} + \lambda \left({\mathfrak p} - {\mathfrak b}\right)
\quad{\text{for}}\quad \lambda\in[0,1]~,
\end{equation}
on the investigated \totem{plex}.
Next, we look at the $I$-divergence of $\mathfrak{l}$ from $\mathfrak u$ as a convex function of $\lambda$,
whose right-derivative at the boundary equals
\begin{equation}
\label{eq:iProj:RightDerivative}
\lim_{\lambda\rightarrow0^+}
\frac{d \infdiv{\mathfrak l}{\mathfrak{u}}}{d \lambda}
= \sum_{\underaccent{\bar}{x}\in\mathcal M}\left(p_x -  b_x \right) \log\frac{b_x}{u_x}
=
\ell(\mathfrak p;\mathfrak b) - \ell(\mathfrak p;\mathfrak{u}) - \infdiv{\mathfrak b}{\mathfrak{u}}
\end{equation}
in terms of the familiar log-likelihood 
\begin{equation}
\label{eq:LogLikelihood}
\ell(\mathfrak p;\mathfrak{u}) = \sum_{\underaccent{\bar}{x}\in\mathcal M} p_x\log u_x
~.
\end{equation}
It is readily recognized that $\ell(\mathfrak p;\mathfrak b)$ diverges as $-\infty$, if any $ b_x$ vanishes whenever $ p_x>0$, while the rest of the metrics remain finite (for compatible reference distribution with $u_x>0$). Hence, the rate of change of the $I$-divergence from $\mathfrak{u}$ while leaving the boundary becomes arbitrarily large and negative. This forces us to conclude that the minimum must correspond to a distribution well in the interior of the \totem{plex} where all of the admissible probabilities are non-vanishing. Thus the  probability vector of the $I$-projection $\mathfrak q \in \mathbb R_{>0}^{\vert\mathcal M\vert}$ has strictly positive entries. 

Having established that the $I$-projection is not a distribution from the boundaries of the \totem{plex} over declared manifestations in Equation~\ref{eq:admissibleEntities}, 
we can apply variational techniques to find its form. Following the standard procedure in optimization theory, we incorporate so-called hard
constraints on expectations 
Equation~\ref{eq:LinearSystem}
via Lagrange multipliers $\boldsymbol\theta$ in the Lagrangian   
\begin{equation}
\label{eq:iProj:Lagrangian}
\mathcal L 
= \infdiv{\mathfrak p}{\mathfrak{u}}
-\boldsymbol\theta^T \left(\mathbf G\:\mathfrak p - \boldsymbol\mu\right)
~.
\end{equation}
Evidently, including some manifestation conditioned by $p_x=0$ would not change $\mathcal L$. Therefore, such manifestations that we have excluded from our \totem{s} in Equation~\ref{eq:ZeroConditions} are not ``seen'' by variational methods and there is no Lagrange multiplier to set $p_x=0$ as a hard constraint.

Subsequently, we extremize the Lagrangian in probabilities $p_x$ and Lagrange multipliers $\boldsymbol \theta\in\mathbb R^{D}$ to directly arrive at the exponential form of $\mathfrak q$ over attribute manifestations in the \totem description:
\begin{equation}
\nonumber
q_x = u_x \exp \left\{\left(\mathbf G^T \boldsymbol\theta\right)_x - 1\right\}
\end{equation}
Using exactly $D=\rank\mathbf G$ Lagrange multipliers (which are eventually fixed to satisfy the non-linear in $\boldsymbol\theta$ system descending from Equation~\ref{eq:LinearSystem}) ensures the independence of the incorporated constraints. This becomes essential for the convergence of many iterative algorithms and important for their efficiency. Due to symmetry Equation~\ref{eq:TOTEM:Reparametrization}, the parametrization in terms of Lagrange multipliers is generically non-unique, whereas the probabilities $q_x$ are due to paramount theorem Equation~\ref{eq:iProjection} unique. 
In other words, only the dual picture in probability space carries physical meaning. 

Since our coefficient matrix $\mathbf G$ shall always imply normalization on the simplex, 
\begin{equation*}
\exists\,\mathbf s\in\mathbb R^D~: \quad \mathbf s^T \mathbf G 
= \mathbf 1^T~,
\end{equation*}
shifting accordingly
\begin{equation*}
\nonumber
\boldsymbol \theta \quad{\rightarrow}\quad \boldsymbol \theta + \mathbf s
\end{equation*}
results into the more compact expression 
\begin{equation}
\label{eq:iProj:parametricForm}
q_x = u_x \exp\left\{\left(\mathbf G^T \boldsymbol\theta\right)_x \right\}  \quad{\text{for}}\quad \underaccent{\bar}{x}\in\mathcal M
~.
\end{equation}
The converse direction also holds: any distribution on $\mathcal T$ of the deduced  form coincides with the $I$-projection of $\mathfrak{u}$ onto the \totem{plex}.
Substituting Equation~\ref{eq:iProj:parametricForm} into the right-hand side of Equation~\ref{eq:Pythagoras} we see that 
\begin{equation}
\infdiv{\mathfrak q}{\mathfrak{u}} + \infdiv{\mathfrak p}{\mathfrak q} = \infdiv{\mathfrak p}{\mathfrak{u}} + 
\boldsymbol\theta^T\underbrace{\mathbf G\left(\mathfrak q -  \mathfrak p\right)}_{=\,0\text{ on }\mathcal T}
\end{equation}
for any probability vector $\mathfrak p\in\mathcal T$.
Hence, the variationally optimal solution Equation~\ref{eq:iProj:parametricForm} fulfills the relation of Pythagorean Lemma and is the unique $I$-projection of $\mathfrak{u}$ onto $\mathcal T$. 

\paragraph{Numerical solution to conditional minimization of information divergence}
In most cases, the minimum of $I$-divergence in the \totem{plex} will not have some closed-form solution expressed in terms of  non-trivial constants $\boldsymbol\mu$.
In this paragraph, we derive the information-projector ($I$-projector) to numerically compute the probabilities of the $I$-projection of $\mathfrak u$ on $\mathcal T$ in the most efficient  way.

Starting from some reference distribution $\mathfrak{u}$, we need to iteratively impose the conditions from Equation~\ref{eq:LinearSystem}, which we summarize in a vector function 
\begin{equation}
\label{eq:Newton:constraintFunctions}
\mathbf v[\mathfrak p] = 
\mathbf G \:\mathfrak p - \boldsymbol\mu
~,
\end{equation}
where $\mathfrak p$ is parameterized according to Equation~\ref{eq:iProj:parametricForm}.
A well-known numerical method to find the root $\mathfrak q$ in the resulting  non-linear system of coupled equations is the Newton-Raphson algorithm, 
which has inspired many iterative methods (listed e.g.\ in~\cite{malouf_comparison_2002}) for approximate inference.
We adapt the Newton-Raphson algorithm to give the covariant (i.e.\ parameter-independent) minimization of $I$-divergence under generic linear conditions Equation~\ref{eq:LinearSystem}. 

Proceeding as usual to find the root of vector-valued function Equation~\ref{eq:Newton:constraintFunctions} given its running estimate $\mathbf v^{(n)} \equiv \mathbf v[\mathfrak p^{(n)}]\in\mathbb R^D$, we update some valid parametrization by 
\begin{equation}
\label{eq:Newton:Parameters_updateRule}
\boldsymbol\theta^{(n+1)} = \boldsymbol\theta^{(n)} - (\mathbf J^{(n)})^{-1}\mathbf v^{(n)}
\end{equation}
in terms of the Jacobian (after taking the derivative of Equation~\ref{eq:Newton:constraintFunctions} w.r.t.\ $\boldsymbol\theta$)
\begin{equation}
\label{eq:NewtonRaphson:Jacobian}
\mathbf J^{(n)} \equiv \mathbf J[\mathbf P^{(n)}] =
\mathbf G\mathbf P^{(n)} \, \mathbf G^T
~.
\end{equation}
To condense notation we use  the diagonal map in $\vert\mathcal M\vert$ dimensions
\begin{equation}
\label{eq:DiagonalMap}
\text{diag} \,: \quad \mathbb R^{\vert\mathcal M\vert}_{>0} \rightarrow \mathbb R^{\vert\mathcal M\vert\times \vert\mathcal M\vert}_{>0} \quad{\text{with}}\quad \mathbf P \equiv \text{diag}(\mathfrak p) 
= 
\begin{pmatrix}
\ddots & & \\[-1ex]
& p_x & \\[-1ex]
& & \ddots
\end{pmatrix} 
~. 
\end{equation}
The matrix form of the update rule for Lagrange multipliers 
\begin{equation}
\mathbf G^T \boldsymbol\theta^{(n+1)} =
\mathbf G^T\boldsymbol\theta^{(n)} - \mathbf G^T (\mathbf J^{(n)})^{-1} \mathbf v^{(n)} \quad
\nonumber
\end{equation} 
implies the update rule for the probability of manifestation $\underaccent{\bar}{x}\in\mathcal M$,
\begin{equation}
u_x \exp\left\{\left(\mathbf G^T \boldsymbol\theta^{(n+1)}\right)_x\right\} = u_x \exp\left\{\left(\mathbf G^T \boldsymbol\theta^{(n)}\right)_x\right\} \exp\left\{-\left(\mathbf G^T (\mathbf J^{(n)})^{-1} \mathbf v^{(n)}\right)_x\right\}
~,
\nonumber
\end{equation}
which we eventually recognize to be
\begin{equation}
 \label{eq:NewtonStep}
p^{(n+1)}_x = p^{(n)}_x \exp\left\{- \left(\mathbf G^T  (\mathbf J^{(n)})^{-1}  \mathbf v^{(n)}\right)_x\right\}~. 
\end{equation} 

Since the Jacobian Equation~\ref{eq:NewtonRaphson:Jacobian} in probability space is expressed in terms of the coefficient matrix only, we manage to remove the dependence on arbitrary parameters in order to write the algorithm exclusively using probabilities of manifestations.
The covariance of algorithm Equation~\ref{eq:NewtonStep} manifests due to the affine invariance of the Newton-Raphson method~\cite{boyd_vandenberghe_2004}. 
Under a physically equivalent re-parametrization of the linear system via $\mathbf S$ in Equation~\ref{eq:TOTEM:Reparametrization}, the condition function Equation~\ref{eq:Newton:constraintFunctions} covariantly transforms as a vector, $\mathbf v \rightarrow \mathbf S\,\mathbf v$,
while the Jacobian  as a matrix, $\mathbf J \rightarrow \mathbf S\,\mathbf J \,\mathbf S^T$. In total, we readily verify that the update rule of Equation~\ref{eq:NewtonStep} remains invariant under re-parametrizations of Equation~\ref{eq:TOTEM:Reparametrization} exemplifying the merits of working in probability space, as opposed to parameter space.
Programmatically, this property streamlines not only the conditional minimization of $I$-divergence in versatile settings, but helps to generate model distributions faster, independently of the current empirical framework.

The iterative algorithm to covariantly  approach the probabilities of the $I$-projection of $\mathfrak{u}$ onto $\mathcal T$
is well-behaved, whenever $\mathfrak q=\argmin_{\mathfrak p\in\mathcal T}\infdiv{\mathfrak p}{\mathfrak{u}}$ is reasonably close to  reference distribution $\mathfrak{u}$.
At the $n$-th step, the $D\times D$ Jacobian $\mathbf J^{(n)}$ evaluated at the running estimate $\mathbf P^{(n)}$ 
is invertible --\,which is a necessary condition for convergence. 
To see this, recall that for any real symmetric positive definite matrix $\mathbf P\succ0$ and $\mathbf G$ of full row-rank with a column-structure compatible with $\mathbf P$, the adjoint map $\mathbf G \mathbf P \mathbf G^T$ from the column to the row space of $\mathbf G$ is also given by a symmetric positive definite and thus invertible matrix:
\begin{align}
\label{eq:Jacobian:PositiveDefinite}
\forall\,\mathbf u \in \mathbb R^{\rank\mathbf G}\,:
0 \overset{!}{=} &\,\,
\mathbf u^T \left(\mathbf G \mathbf P \mathbf G^T\right)\mathbf u
=
\left(\mathbf G^T\mathbf u\right)^T \mathbf P \left(\mathbf G^T\mathbf u\right)
&\overset{\mathbf P\succ0}{\Longrightarrow} \quad
~
\mathbf u^T\mathbf G = \mathbf 0 
\quad{\overset{\rank\mathbf G=D}{\Longrightarrow}}\quad
\mathbf u = \mathbf 0
~.
\end{align}
After $n$ iterations,  we recognize by the scalar product
\begin{equation}
\left(\boldsymbol\theta^{(n+1)} - \boldsymbol\theta^{(n)}\right)^T \mathbf v^{(n)}
=
- \mathbf v^{(n)} (\mathbf J^{(n)})^{-1} \mathbf v^{(n)} < 0 \quad\forall \,n\quad\text{where\ }\quad \mathbf v^{(n)} \neq \mathbf 0
\nonumber
\end{equation}
using Equation~\ref{eq:Newton:Parameters_updateRule} and positive-definiteness of $(\mathbf J^{(n)})^{-1}$ that the procedure described by Equation~\ref{eq:NewtonStep} moves along a descent direction until the desired convergence.

If some probabilities become exponentially suppressed over the optimization cycles (hence $\mathbf P^{(n)}$ tends to a diagonal \textit{semi}-positive definite matrix), this hints towards manifestations whose probability is forced to vanish by the \totem conditions, thus violating defining Equation~\ref{eq:ZeroConditions}. In that case, one has to rerun the iterative optimizer after redefining the \totem{plex} over admissible manifestations only, which are allowed by Equation~\ref{eq:LinearSystem} to attain non-vanishing $p_x$.

\paragraph*{Successive $I$-projection{s} in nested descriptions}
Since the Newton-Raphson estimate in Equation~\ref{eq:NewtonStep} remains agnostic to the concrete parametrization of Equation~\ref{eq:iProj:parametricForm}, convergence of the iterative procedure solely depends on the physical setup. If the reference distribution $\mathfrak{u}$ poorly captures the expectations described by $\boldsymbol \mu$
resulting into higher $I$-divergence  $\infdiv{\mathfrak q}{\mathfrak{u}}$, the algorithm might fail to converge to the $I$-projection $\mathfrak q$, as generally expected by Newton-based methods~\cite{boyd_vandenberghe_2004}. 
In that case, we can try to successively impose the conditions in Equation~\ref{eq:LinearSystem}, to gradually move away from the reference distribution by calculating a chain of $I$-projections on nested \totem{plices}. 

For nested \totem{s} $\mathcal T_B\subset\mathcal T_A$ in the sense of defining Equation~\ref{eq:NestedTOTEMs:CoefficientMatrix} and Equation~\ref{eq:NestedTOTEMs:moments}, the $I$-projection of $\mathfrak{u}$ onto the inner \totem{plex} satisfies the chain rule
\begin{equation}
\label{eq:iProj:ChainRule}
\mathfrak q_B = \argmin_{\mathfrak p\in\mathcal T_B} \infdiv{\mathfrak p}{\mathfrak{u}}
=
\argmin_{\mathfrak p\in\mathcal T_B} \infdiv{\mathfrak p}{\mathfrak q_A} \quad{\text{with}}\quad \mathfrak q_A = \argmin_{\mathfrak p\in\mathcal T_A}\infdiv{\mathfrak p}{\mathfrak{u}}
\end{equation}
coinciding with the $I$-projection onto the inner \totem{plex} $\mathcal T_B$ of the $I$-projection $\mathfrak q_A$ of $\mathfrak{u}$ onto the ambient \totem{plex} $\mathcal T_A$. To see this we start from the definition of the $I$-projection of $\mathfrak q_A$ onto the inner $\mathcal T_B$, which we call $\mathfrak q_B'$:
\begin{equation}
\label{eq:Newton:nestedTOTEMs:Proof}
\forall\,\,{\mathfrak p}\in \mathcal T_B\, : \quad 
\infdiv{\mathfrak p}{\mathfrak q_A}
\geq 
\infdiv{ \mathfrak q_B' }{ \mathfrak q_A }   \quad{\text{with}}\quad \mathfrak q_B' \equiv \argmin_{\mathfrak p\in\mathcal T_B} \infdiv{\mathfrak p}{\mathfrak q_A}
~.
\end{equation}
Substituting twice the Pythagorean Lemma in Equation~\ref{eq:Pythagoras} (recall that $\mathfrak q_A$ itself is the $I$-projection of $\mathfrak{u}$ on ambient $\mathcal T_A$) applied for distributions $\mathfrak p,\mathfrak q_B'\in\mathcal T_A$,
\begin{align}
\infdiv{\mathfrak p}{\mathfrak q_A} =
\infdiv{\mathfrak p}{\mathfrak{u}} - 
\infdiv{\mathfrak q_A}{\mathfrak{u}}
\quad\text{and}\quad
\nonumber
\infdiv{\mathfrak q_B'}{\mathfrak q_A}
=
\infdiv{\mathfrak q_B'}{\mathfrak{u}} - \infdiv{\mathfrak q_A}{\mathfrak{u}}
\nonumber
~,
\end{align}
cancels the common term $\infdiv{\mathfrak q_A}{\mathfrak{u}}$ on both sides of the inequality in Equation~\ref{eq:Newton:nestedTOTEMs:Proof} leading to 
\begin{equation}
\infdiv{\mathfrak p}{\mathfrak{u}} 
\geq 
\infdiv{\mathfrak q_B'}{\mathfrak u}
~.
\end{equation}
Since this applies for all $\mathfrak p$ in the inner \totem{plex}, we conclude by uniqueness of the $I$-projection of $\mathfrak u$ on $\mathcal T_B$ that the distribution  $\mathfrak q_B'$ (introduced as the $I$-projection of $\mathfrak q_A$ onto $\mathcal T_B$) is also the $I$-projection of $\mathfrak{u}$ onto $\mathcal T_B$ meaning $\mathfrak q_B'=\mathfrak q_B$.

Let us suppose that the original problem requires to compute the $I$-projection of $\mathfrak{u}$ on $\mathcal T_B$ but the $I$-projector fails to converge. In that case, we might try to first compute the $I$-projection of $\mathfrak{u}$ on some ambient $\mathcal T_A$ incorporating only a subset of the original conditions such that $\mathcal T_B\subset\mathcal T_A$. By the Pythagorean property $\infdiv{\mathfrak q_B}{\mathfrak{u}} = \infdiv{\mathfrak q_B}{\mathfrak q_A} + \infdiv{\mathfrak q_A}{\mathfrak{u}}$ of the $I$-projection of $\mathfrak{u}$ on ambient $\mathcal T_A$, it is 
\begin{equation*}
\infdiv{\mathfrak q_A}{\mathfrak{u}} \leq \infdiv{\mathfrak q_B}{\mathfrak{u}}~.
\end{equation*}
Therefore, the $I$-projector for $\mathfrak q_A = \argmin_{\mathfrak p\in\mathcal T_A}\infdiv{\mathfrak p}{\mathfrak{u}}$ becomes more likely to converge than directly projecting from $\mathfrak{u}$ on $\mathcal T_B$. Subsequently, we compute the $I$-projection of $\mathfrak q_A$ on $\mathcal T_B$, the \totem{plex} defined by all the original conditions. Similarly, the $I$-projector is more likely to converge due to 
\begin{equation*}
\infdiv{\mathfrak q_B}{\mathfrak q_A} \leq \infdiv{\mathfrak q_B}{\mathfrak{u}}~.
\end{equation*}
By the chain rule proved in Equation~\ref{eq:iProj:ChainRule}, the successive application of the $I$-projector using one intermediate \totem{plex} $\mathcal T_A$ gives the desired $I$-projection of $\mathfrak{u}$ directly on $\mathcal T_B$. If necessary to achieve convergence within desired accuracy, the generalization to more intermediate $I$-projections on nested \totem{plices} is straightforward.

\paragraph{The large-$N$ expansion around the $I$-projection}
In the universality class of multinomial-type sampling conditioned on the \totem{plex} $\mathcal T$, the large-$N$ approximation  to the sampling probability in Equation~\ref{eq:MutlinomialSampling:LargeN} reveals a spherical symmetry in $p_x$ space. This realization, which is purely based on the large-$N$ asymptotics of quasi-independent sampling processes, leads in turn to the \totem $\chi^2$ statistic. 

Exploiting the Pythagorean Equation~\ref{eq:Pythagoras} for $I$-divergences, we can rewrite the sampling probability $\text{Pr}(N\mathfrak p)\equiv \text{Pr}(N\mathfrak p\vert\mathcal T;\mathfrak u)$ in Equation~\ref{eq:MutlinomialSampling:LargeN} as 
\begin{equation}
\label{eq:MutlinomialSampling:LargeN_reDefined}
\log\text{Pr}(N\mathfrak p) 
=
- N \infdiv{\mathfrak p}{\mathfrak q} 
\, + \, \mathcal O(1) - \log  Z
\quad,\quad
\mathfrak p \in \mathcal T
\end{equation}
by absorbing constants into redefined partition sum $Z$. 
The dependence on the reference distribution $\mathfrak{u}$ (which does not need to belong to the \totem{plex}) is implicit through $\mathfrak q$, the $I$-projection of $\mathfrak{u}$ on $\mathcal T$. The latter also carries the dependence on $\boldsymbol \mu$, according to the linear system in Equation~\ref{eq:LinearSystem} which defines the \totem{plex}.

Next, we parameterize distributions $\mathfrak p\in\mathcal T$ by probability fluctuations $\boldsymbol\psi\in\ker\mathbf G$  
around $\mathfrak q$. Stochastic fluctuations span the appropriate polyhedral cone such that the  distribution 
\begin{equation}
\label{eq:Fluctuations}
\mathfrak p = \mathfrak q + \frac{\boldsymbol\psi}{\sqrt{N}} \quad\text{with}\quad \mathfrak q = \argmin_{\mathfrak p\in\mathcal T} \infdiv{\mathfrak p}{\mathfrak{u}}
\end{equation}
remains within $\mathcal T$. By the linear parametrization, stochastic fluctuations satisfy a homogeneous system 
\begin{equation}
\mathbf G \, \boldsymbol \psi = \boldsymbol 0~,
\end{equation}
in particular $\text{Tr}\,\mathbf \Psi=0$ with $\mathbf\Psi=\text{diag}\,\boldsymbol\psi$ (see Equation~\ref{eq:DiagonalMap}) due to normalization on the simplex.
Therefore, the large-$N$ expansion of sampling distribution in Equation~\ref{eq:MutlinomialSampling:LargeN_reDefined} around the $I$-projection becomes
\begin{equation}
\label{eq:MutlinomialSampling:Gaussianization}
\log\text{Pr}(N\mathfrak p) = 
-\tfrac12 \boldsymbol\psi^T\mathbf Q^{-1}\boldsymbol\psi
+ \mathcal O(N^{-1/2}) - \log Z
\quad\text{with}\quad \mathbf Q = \text{diag}\,\mathfrak q
~.
\end{equation}
The absence of a linear term in the fluctuations verifies that we indeed expand around the global minimum of Equation~\ref{eq:iProjection}.

When $N$ is sufficiently large, standard Laplace approximation around the unique saddle point (i.e.\ the $I$-projection) allows us to extend the limits of integration to infinity, so that Equation~\ref{eq:MutlinomialSampling:Gaussianization} becomes a $\vert\ker\mathbf G\vert$-variate normal distribution. It is then easy to recognize that any higher-order term in the fluctuations $\boldsymbol \psi$ is suppressed by inverse powers of $\sqrt{N}$. The normalization factor $Z$ is fixed by self-consistency of the large $N$-expansion, meaning the conditional sampling probability on the \totem{plex} must sum to unity. $N$-subleading corrections can be treated perturbatively, i.e.\ integrated out order by  order using the $\vert\ker\mathbf G\vert$-variate Gaussian density.

Indeed, $\boldsymbol\psi$ can be expressed through a vector $\boldsymbol\varphi\in\mathbb R^{\vert\ker\mathbf G\vert}$ along 
the orthogonal complement $\mathbf K$ (the so-called null space with $\text{span}\,\mathbf K=\ker\mathbf G$) to the space spanned by $\mathbf G$,
\begin{equation}
\label{eq:Fluctuations:OrthogonalComplement}
\boldsymbol\psi = 
\mathbf K^T\boldsymbol\varphi
\quad\text{for}\quad \mathbf G\,\mathbf K^T = \boldsymbol 0~.
\end{equation}
Comparing Equation~\ref{eq:MutlinomialSampling:LargeN_reDefined} to Equation~\ref{eq:MutlinomialSampling:Gaussianization} in the continuum limit of large-$N$ asymptotics,  
\begin{equation}
\text{Pr}(N\mathfrak p)\quad\rightarrow\quad
\rho(\boldsymbol\psi)\equiv \rho(\boldsymbol\varphi)~,
\end{equation}
we observe that the quadratic form of the multivariate Gaussian density $\rho$,
\begin{equation}
\label{eq:QuadraticForm}
0\leq
N\infdiv{\mathfrak p}{\mathfrak q} \approx 
\tfrac12 \boldsymbol\psi^T\mathbf Q^{-1}\boldsymbol\psi
= 
\tfrac12\boldsymbol\varphi^T \mathbf K \,\mathbf Q^{-1}\mathbf K^T \boldsymbol\varphi
~,
\end{equation}
reflects the spherical symmetry of information divergence in the kernel space of linear system Equation~\ref{eq:LinearSystem}. 
Due to Equation~\ref{eq:Jacobian:PositiveDefinite}, the precision matrix $\boldsymbol\Sigma^{-1}\equiv\mathbf K\, \mathbf Q^{-1}\mathbf K^T$ (recall that $\mathbf Q=\text{diag}\,\mathfrak q\succ 0$ by Equation~\ref{eq:iProj:parametricForm}) of the quadratic form is positive definite, confirming the stability of Laplace approximation via a well-behaved Gaussian density $\rho$.

\paragraph{The emergence of $\chi^2$ statistic}
The ``Gaussianization'' achieved in Equation~\ref{eq:MutlinomialSampling:Gaussianization} of the conditional sampling probability in the \totem description gives rise to a spherical symmetry in $p_x$ space.
Any distribution $\mathfrak p$ on the \totem{plex} is sampled from ${\mathfrak{u}}$ as a kernel fluctuation $\boldsymbol\varphi\equiv\boldsymbol\varphi(\mathfrak p)$ away from the $I$-projection $\mathfrak q$ 
using the $N$-leading probability 
\begin{equation}
\label{eq:ProbabilityToSample}
\rho(\boldsymbol\varphi) \prod_{j=1}^{k} d \varphi_j
\quad\text{with}\quad \rho(\boldsymbol\varphi) = Z^{-1} \exp\left\{-\tfrac 12
\boldsymbol\varphi^T \mathbf K \,\mathbf Q^{-1}\mathbf K^T \boldsymbol\varphi
\right\}~.
\end{equation}
in the $k\equiv\vert\ker\mathbf G\vert$-dimensional space.

Let $\lambda_j>0$ denote the $j$-th eigenvalue of the quadratic form on the right-hand side of Equation~\ref{eq:QuadraticForm}  
and group its orthonormal eigenvectors into the columns of a $k\times k$ orthogonal matrix $\mathbf U$.
Setting $\boldsymbol\varphi =  \mathbf U \boldsymbol x$ we diagonalize Equation~\ref{eq:QuadraticForm} in the canonical fluctuations $\boldsymbol x\in\mathbb R^{k}$ with unit-$\det$ Jacobian, so that the quadratic form becomes $\sum_{j=1}^k \lambda_j x_j^2$.
Subsequently, we eliminate the positive eigenvalues by $y_j = \sqrt{\lambda_j}\, x_j$ where the Jacobian $\prod_j \lambda_j^{-1/2}=\mathcal O(1)$ of the total differential in Equation~\ref{eq:ProbabilityToSample} gets canceled  by the partition function $Z$. 
Switching eventually to spherical coordinates with unit-$\det$ Jacobian via $r^2 = \sum_{j=1}^k y_j^2$, the probability in Equation~\ref{eq:ProbabilityToSample} becomes
\begin{equation}
\label{eq:ProbabilityToSample:ShpericalMeasure}
d\Omega_k\, d r\,  r^{k-1} \rho(r)~,
\end{equation}
with radial density
\begin{equation}
\rho(\boldsymbol\varphi) \equiv \rho(r) = Z^{-1}\exp \left\{ -\tfrac12 r^2\right\}~.
\end{equation}
It is trivial to execute the angular integration, where the solid angle in $k$ dimensions equals 
\begin{equation}
\Omega_k = \frac{2\pi^{k/2}}{\Gamma(k/2)}~.
\end{equation}

Once again, we change integration variable to 
\begin{equation}
\label{eq:ChiSquaredStatistics}
Q= r^2  = 2 N \infdiv{\mathfrak p}{\mathfrak q}
\end{equation}
in order to connect with more established conventions in the statistics literature. The second equality refers back to Equation~\ref{eq:QuadraticForm}. 
In the squared radius $Q$, the cumulative conditional probability on the $(k-1)$-sphere from Equation~\ref{eq:ProbabilityToSample:ShpericalMeasure} reads
\begin{equation}
\label{eq:TOTEMplex:ChiSquaredDistro}
d Q \,
\frac{1}{2^{k/2}\Gamma(k/2)}Q^{k/2-1}\, e^{-Q/2}
\quad\text{for}\quad Q\geq0
~.
\end{equation}
This expresses the conditional probability at $N\gg1$  to sample any dataset $N\mathfrak p$ whose $I$-divergence from the $I$-projection lies on a spherical shell of radius $\frac{Q}{2N}$ and thickness $d Q$. We readily recognize that the $N$-leading approximation to the conditional sampling probability follows a $\chi^2$ distribution with $k=\vert\ker\mathbf G\vert=\vert\mathcal M\vert-D$ degrees of freedom and statistic given by Equation~\ref{eq:ChiSquaredStatistics}. 
Since $Q$ is $\chi^2$ distributed, its variance is $\mathcal O(1)$ in the large-$N$ expansion, so that the variance of the $I$-divergence $\infdiv{\mathfrak p}{\mathfrak q}$ dominating the sampling probability of $\mathfrak p$ in Equation~\ref{eq:MutlinomialSampling:LargeN_reDefined} scales as $\mathcal O(1/N)$. This shows that the larger $N$ becomes, the more probability mass on the \totem{plex} is ``concentrated'' around $\mathfrak q$, the $I$-projection of $\mathfrak{u}$ on $\mathcal T$.

If the uniform distribution $\mathfrak{u}_\text{u}=\vert\mathcal M\vert^{-1}\left(1,\ldots,1\right)^T$ is used as a reference, when $I$-divergence minimization reduces to entropy maximization, 
\begin{equation}
\label{eq:Idiv_Entropy:Relation}
\infdiv{\mathfrak p}{\mathfrak u_\text{u}} = -H[\mathfrak p] + \log \vert\mathcal M\vert~,
\end{equation}
the derived $\chi^2$ distribution reduces to the so-called concentration theorem of entropies derived in~\cite{Rosenkrantz1989}. Albeit excessively motivated in the context of prior analysis for Bayesian inference, there is no appealing (empirical or mathematical) motivation  to insist on the uniform distribution as a reference. 

As a further connection to the model-centric literature which exemplifies the unifying character of \totem, we reconsider the radial density in Equation~\ref{eq:ProbabilityToSample:ShpericalMeasure},
\begin{equation}
\tilde \rho(r) \equiv r^{k-1} \rho(r) ~,
\end{equation}
where the conditional probability density $\rho$ from Equation~\ref{eq:ProbabilityToSample} is expressed in spherical coordinates. At some $\mathfrak p\in\mathcal T$, for which Equation~\ref{eq:ChiSquaredStatistics} prescribes $r\equiv r(\mathfrak p)$, the logarithm of the redefined density introduces the 
\begin{equation}
\text{score}_I = \log\tilde\rho(r) = - N \infdiv{\mathfrak p}{\mathfrak q} + \frac{k-1}{2} \log N + \mathcal O(1)~.
\end{equation}
Setting $\mathfrak p=\mathfrak f$, this score immediately reproduces the well-known Bayesian information criterion~\cite{loukas_total_2023}, 
\begin{equation}
    \textsc{bic} = -2N\ell(\mathfrak f;\mathfrak q) + D \log N = \left(-2\right) \cdot \text{score}_I + \ldots
\end{equation}
up to irrelevant constants (given data and manifestations) which modify the overall normalization. $\ell$ is the log-likelihood defined in Equation~\ref{eq:LogLikelihood}, $\mathfrak q$ gives the information-theoretic counterpart of  a ``model'' and $D=\rank\mathbf G = \vert\mathcal M\vert - k$ alludes to the dimension of the ``model''. Given a candidate list of conditions $\mathbf G$ and control vectors $\boldsymbol\mu$, the principle of maximum likelihood would dictate to select the candidate \totem system whose associated $I$-projection maximizes the $\text{score}_I$; equivalently minimizing the \textsc{bic}.

\paragraph{The information test}
Finally, we use the conditional sampling probability in Equation~\ref{eq:MutlinomialSampling:LargeN_reDefined} to formulate statistical hypothesis testing in a fully empirical manner. For that, we exploit the ``concentration'' of probability mass on the \totem{plex} around the $I$-projection at large $N$ 

Consider two nested \totem{plices} $\mathcal T_\text{J}\subset\mathcal T_\text{I}$ where $\mathbf G_\text{J}$ implies $\mathbf G_\text{I}$ according to defining Equation~\ref{eq:NestedTOTEMs:CoefficientMatrix}, i.e.\ $\mathbf G_\text{I} = \mathbf S\, \mathbf G_\text{J}$ meaning\footnote{Let $\boldsymbol n\in\ker\mathbf G_\text{J}$, i.e.\ $\mathbf G_\text{J}\boldsymbol n =\boldsymbol0$. Then nestedness implies also $\mathbf G_\text{I}\boldsymbol n= \mathbf S \left(\mathbf G_\text{J}\boldsymbol n\right) = \boldsymbol 0$. Hence, $\boldsymbol n\in\ker\mathbf G_\text{I}$ as well.} $\ker\mathbf G_\text{J}\subset\ker\mathbf G_\text{I}$, alongside compatible conditions $\boldsymbol \mu_\text{I} = \mathbf S\,\boldsymbol \mu_\text{J}$ in the sense of Equation~\ref{eq:NestedTOTEMs:moments}. Clearly, it is $D_\text{J} > D_\text{I}$ for the row-ranks of $\mathbf G_\text{J}$ and $\mathbf G_\text{I}$, respectively. 
Let us estimate the  probability mass of the inner \totem{plex}, i.e.\ the cumulative sampling probability to sample any distribution on $\mathcal T_\text{J}$, 
\begin{equation}
\text{Pr}(\mathcal T_\text{J} \vert \mathcal T_\text{I}) := \int_{\mathfrak p\in\mathcal T_\text{J}} \rho(\mathfrak p\vert \mathcal T_\text{I})
\end{equation} 
given the large-$N$ conditional probability density $\rho(\mathfrak p\vert \mathcal T_\text{I})$ from Equation~\ref{eq:ProbabilityToSample} on the ambient \totem{plex}. 

First, we examine the ambient \totem{plex} $\mathcal T_\text{I}$ defined by $\mathbf G_\text{I}$ under henceforth fixed conditions $\boldsymbol \mu_\text{I}\in\mathbb R^{D_\text{I}}$.
Stochastic fluctuations Equation~\ref{eq:Fluctuations} around the $I$-projection of $\mathfrak{u}$ on $\mathcal T_\text{I}$ --\,denoted in the following by $\mathfrak q_\text{I}$\,-- decompose by completeness of vector space
into
\begin{equation}
\label{eq:QuotientSpace:Decomposition}
\boldsymbol \psi_\text{I}
=
\mathbf K_\text{I}^T \boldsymbol\varphi_\text{I}  = \mathbf K_n^T \boldsymbol\phi_n
+ 
 \mathbf K_\text{J}^T \boldsymbol\varphi_\text{J}
 ~.
\end{equation}
As before, $\mathbf K_\text{I}$ spans the ambient $\ker\mathbf G_\text{I}$, $\mathbf K_\text{J}$ analogously spans $\ker\mathbf G_\text{J}$ and $\mathbf K_n$ spans the space complementing $\ker\mathbf G_\text{J}$ to $\ker\mathbf G_\text{I}$ with $\rank\mathbf K_n = \vert{\ker \mathbf G_\text{I}}\vert - \vert{\ker \mathbf G_\text{J}}\vert = D_\text{J} - D_\text{I}$.
In fact, this decomposition parameterizes quotient spaces on the simplex. 
The set of $\mathbf G_\text{J}$-induced equivalence classes constitutes namely the quotient space --\,signified by $\mathcal T_\text{I}\slash\mathbf G_\text{J}$\,-- of the ambient \totem{plex} under the equivalence relation
\begin{equation}
\label{eq:QuotientSpace:EquivalenceRelation}
\mathfrak p_1, \mathfrak p_2 \in \mathcal T_\text{I}~:\quad
\mathfrak p_1 \sim \mathfrak p_2 \quad\Longleftrightarrow\quad
\mathbf G_\text{J}\left(\mathfrak p_1-\mathfrak p_2\right) = \boldsymbol 0~.
\end{equation}
The first part of decomposition $\ker\mathbf G_\text{I} =  \mathcal T_\text{I}/\mathbf G_\text{J} \oplus \ker\mathbf G_\text{J}$ in Equation~\ref{eq:QuotientSpace:Decomposition} describes fluctuations relating the $\mathbf G_\text{J}$-distinct equivalence classes on the ambient $\mathcal T_\text{I}$ whereas the second part corresponds to the fluctuations within each $\mathbf G_\text{J}$-induced class at given $\boldsymbol\mu_\text{J}$ values.

Plugging Equation~\ref{eq:QuotientSpace:Decomposition} in $N$-leading Gaussian density $ \rho(\boldsymbol\varphi_\text{I}\vert\mathcal T_\text{I}) \equiv \rho(\boldsymbol\phi_n, \boldsymbol\varphi_\text{J}\vert\mathcal T_\text{I})$ derived in Equation~\ref{eq:MutlinomialSampling:Gaussianization}, 
\begin{align}
\label{eq:NestedTOTEMs:GaussianKernel}
-\log\text{Pr}(N\mathfrak p\vert\mathcal T_\text{I}) + \ldots =&\,\,  N \infdiv{\mathfrak p}{\mathfrak q_\text{I}} 
= 
\tfrac12\boldsymbol\varphi_\text{I}^T
\mathbf \Sigma_\text{I}^{-1} 
\boldsymbol\varphi_\text{I}
=
\tfrac12 \boldsymbol\phi_n^T  \mathbf \Sigma^{-1}_n \boldsymbol\phi_n 
+ 
\tfrac12 \boldsymbol\varphi_\text{J}^T  \mathbf \Sigma_\text{J}^{-1} \boldsymbol\varphi_\text{J}
+
\boldsymbol\varphi_\text{J}^T \mathbf M \boldsymbol\phi_n
\end{align}
breaks the precision matrix on the ambient $\mathcal T_\text{I}$  w.r.t.\ quotient-space decomposition 
\begin{equation}
\boldsymbol\varphi_\text{I} \equiv 
\begin{pmatrix}
\boldsymbol\phi_n \\ 
\boldsymbol\varphi_\text{J}
\end{pmatrix}
\nonumber
\end{equation}
into so-called conformable partitions
\begin{equation}
\mathbf \Sigma_\text{I}^{-1} \equiv 
\begin{pmatrix}
\mathbf \Sigma_n^{-1} & \mathbf M^T\\
\mathbf M & \mathbf \Sigma_\text{J}^{-1}
\end{pmatrix}
\nonumber
\end{equation}
with precision matrices
\begin{equation}
\mathbf \Sigma^{-1}_n = \mathbf K_n\mathbf Q^{-1} {\mathbf K_n}^T
\quad\text{and}\quad
\mathbf \Sigma_\text{J}^{-1} = \mathbf K_\text{J}\mathbf Q^{-1} \mathbf K_\text{J}^T
\nonumber
\end{equation}
and mixed term
\begin{equation}
\mathbf M  = \mathbf K_\text{J}\mathbf Q^{-1} \mathbf K_n^T ~.
\nonumber
\end{equation}
Again by Equation~\ref{eq:Jacobian:PositiveDefinite}, we recognize that fluctuations $\boldsymbol\varphi_\text{J}$ parametrizing $\ker\mathbf G_\text{J}$ and thus sharing same  statistics $\boldsymbol\mu_\text{J}$ are controlled (at large $N$) by a $\vert{\ker\mathbf G_\text{J}}\vert$-variate normal distribution. 

Integrating out $\boldsymbol\varphi_\text{J}$ we are left with a $(D_\text{J} - D_\text{I})$-variate normal distribution (due to closure of Gaussian integrals under marginalization) in $\boldsymbol\phi_n$ that parametrizes the quotient space $\mathcal T_\text{I}/\mathbf G_\text{J}$:
\begin{equation}
\label{eq:nestedTOTEMs:Fluctuations:MarginalizingGaussian}
\rho(\mathcal T_B\vert\mathcal T_\text{I}) 
= 
\int d\boldsymbol\varphi_\text{J}\: \rho(\boldsymbol\phi_n,\boldsymbol\varphi_\text{J}\vert\mathcal T_\text{I}) 
=
\left[\det\left(2\pi \mathbf\Sigma_\text{eff}\right)\right]^{-1/2}
\exp\left\{-\tfrac12\boldsymbol\phi_n^T
\mathbf\Sigma_\text{eff}^{-1}\boldsymbol\phi_n\right\}
\end{equation}
with effective precision matrix
\begin{equation}
\mathbf \Sigma_\text{eff}^{-1}\equiv \mathbf \Sigma^{-1}_n - \mathbf M^T \mathbf\Sigma_\text{J}\mathbf M~.
\nonumber
\end{equation}
Incidentally, the resulting precision matrix $\mathbf\Sigma_\text{eff}^{-1} = \mathbf\Sigma^{-1}_\text{I}/\boldsymbol\Sigma_\text{J}^{-1}$ is called the Schur complement of block $\boldsymbol\Sigma_\text{J}^{-1}$ for which the relation
$\det \mathbf \Sigma^{-1}_\text{I} = \det \boldsymbol\Sigma_\text{J}^{-1} \cdot\det \boldsymbol\Sigma_\text{eff}^{-1}$ holds by the determinant rule. This fixes the normalization of the marginal normal distribution.  
That the precision matrix of latter marginal distribution is positive definite can be easily seen by completing the square for $\boldsymbol\varphi_\text{J}$ in  Equation~\ref{eq:NestedTOTEMs:GaussianKernel}:
\begin{equation}
\label{eq:Fluctuations:NestedTOTEMs:CompleteTheSquare}
\boldsymbol\varphi_\text{I}^T \mathbf \Sigma^{-1}_\text{I}\boldsymbol\varphi_\text{I} = \left(\boldsymbol\varphi_\text{J} + \mathbf \Sigma_\text{J}\mathbf M\boldsymbol\phi_n \right)^T \mathbf\Sigma_\text{J}^{-1} \left(\boldsymbol\varphi_\text{J} + \mathbf \Sigma_\text{J}\mathbf M\boldsymbol\phi_n \right) + \boldsymbol\phi_n^T \boldsymbol\Sigma^{-1}_\text{eff}\boldsymbol\phi_n
\end{equation}
Per assumption of independent fluctuations in the ambient kernel Equation~\ref{eq:Fluctuations:OrthogonalComplement}, the left-hand side always stays positive for any $\boldsymbol\varphi_\text{I}\neq\boldsymbol 0$, thus for any vector $\boldsymbol\varphi_\text{J}\neq\boldsymbol 0$ and $\boldsymbol\phi_n\neq\boldsymbol 0$.
Choose in particular $\boldsymbol\varphi_\text{J} = -\boldsymbol \Sigma_\text{J}\mathbf M\boldsymbol\phi_n$ so that 
\begin{equation}
\forall\,\boldsymbol\phi_n \in \mathbb R^{D_\text{J}-D_\text{I}} \setminus \{\boldsymbol 0\}\, :\quad 0 < \boldsymbol\varphi_\text{I}^T
\mathbf \Sigma^{-1}_\text{I} \boldsymbol\varphi_\text{I} = \boldsymbol\phi_n^T \boldsymbol\Sigma^{-1}_\text{eff} \boldsymbol\phi_n
~.
\end{equation}
Hence, $ \boldsymbol\Sigma^{-1}_\text{eff}$ has to be positive definite. A derivation analogous to the one leading to Equation~\ref{eq:TOTEMplex:ChiSquaredDistro}, the ``concentration of probablity'', via spherical symmetry shows that the marginal Gaussian density in Equation~\ref{eq:nestedTOTEMs:Fluctuations:MarginalizingGaussian} expresses a $\chi^2$ distribution with $(D_\text{J}-D_\text{I})$ degrees of freedom.

By the Gaussian integration over $\boldsymbol\varphi_\text{J}$ in Equation~\ref{eq:nestedTOTEMs:Fluctuations:MarginalizingGaussian}, the resulting quadratic form is nothing but the minimum of the ambient quadratic form $\boldsymbol\varphi_\text{I}^T \mathbf \Sigma^{-1}_\text{I}\boldsymbol\varphi_\text{I}$ in Equation~\ref{eq:Fluctuations:NestedTOTEMs:CompleteTheSquare} when viewed as a function of $\boldsymbol\varphi_\text{J}$ at fixed $\boldsymbol\phi_n$.
Since $\boldsymbol\varphi_\text{J}$ parametrizes $\ker\mathbf G_\text{J}$, i.e.\ a nested \totem{plex} $\mathcal T_\text{J}$ with $I$-projection $\mathfrak q_\text{J}$ in the quotient-space decomposition of Equation~\ref{eq:QuotientSpace:Decomposition}, minimizing the ambient quadratic form amounts to 
\begin{equation}
    \min_{\boldsymbol\varphi_\text{J}} \tfrac12\boldsymbol\varphi_\text{I}^T \mathbf \Sigma^{-1}_\text{I}\boldsymbol\varphi_\text{I} = \min_{\mathfrak p\in\mathcal T_\text{J}} N\infdiv{\mathfrak p}{\mathfrak q_\text{I}} = N\infdiv{\mathfrak q_\text{J}}{\mathfrak q_\text{I}}
\end{equation}
using Equation~\ref{eq:NestedTOTEMs:GaussianKernel}.
By the chain rule of Equation~\ref{eq:iProj:ChainRule}, the $I$-projection of $\mathfrak q_\text{I}$ on the inner \totem{plex} is equivalent  to
\begin{equation}
\mathfrak q_\text{J} = \argmin_{\mathfrak p\in\mathcal T_\text{J}}\infdiv{\mathfrak p}{\mathfrak q_\text{I}}\overset{}{=} \argmin_{\mathfrak p\in\mathcal T_\text{J}}\infdiv{\mathfrak p}{\mathfrak{u}}
~.
\end{equation} 
Putting everything together, we obtain the main result of this paragraph (dropping $N$-subleading corrections):
\begin{equation}
\label{eq:NestedTOTEMs:MaximumOfGaussianMarginalization}
\log\int d\boldsymbol\varphi_\text{J}\: \rho(\boldsymbol\phi_n,\boldsymbol\varphi_\text{J}\vert\mathcal T_\text{I})  = 
-\tfrac12\boldsymbol\phi_n^T\mathbf\Sigma_\text{eff}^{-1}\boldsymbol\phi_n 
=
-\tfrac12\min_{\boldsymbol\varphi_\text{J}} \boldsymbol\varphi_\text{I}^T \mathbf \Sigma^{-1}_\text{I}\boldsymbol\varphi_\text{I} 
= -N\infdiv{\mathfrak q_\text{J}}{\mathfrak q_\text{I}} ~.
\end{equation}
From this expression, we recognize that all \totem{plices} $\mathcal T_{\text{J}'}\subset\mathcal T_\text{I}$ constructed by $\mathbf G_\text{J}$ with $\boldsymbol\mu_\text{I}=\mathbf S\boldsymbol\mu_{\text{J}'}$ that exhibit the same minimal $I$-divergence from $\mathfrak u$ (the $I$-divergence of the $I$-projection of $\mathfrak u$ on $\mathcal T_{\text{J}'}$) have approximately the same sampling probability.

Consequently, the spherical symmetry arises this time in the quotient space $\mathcal T_\text{I}/\mathbf G_\text{J}$ parameterized by normal fluctuations $\boldsymbol\phi_n$ and concerns the sampling probability of whole \totem{plices}, i.e.\ the cumulative probability to select any distribution on $\mathcal T_{\text{J}'}\subset\mathcal T_\text{I}$. This establishes that the conditional cumulative probability to sample the nested $\mathcal T_\text{J}\subset\mathcal T_\text{I}$ at the given values $\boldsymbol\mu_\text{J}$ or any other nested \totem{plex} exhibiting a minimal $I$-divergence of $\infdiv{\mathfrak q_\text{J}}{\mathfrak q_\text{I}}$ is approximated at large $N$ by 
\begin{equation}
\label{eq:iTest:ChiSquaredStatistics}
\rho(\mathcal T_\text{J}\vert\mathcal T_\text{I}) =\chi^2_{D_\text{J}-D_\text{I}}(2 N \infdiv{\mathfrak q_\text{J}}{\mathfrak q_\text{I}}) \quad\text{with}\quad \mathfrak q_{i} = \argmin_{\mathfrak p\in\mathcal T_{i}}  \infdiv{\mathfrak p}{\mathfrak{u}} \quad\text{for}\quad i=\text{I}, \text{J}~.
\end{equation}
In that way, we can compare the likelihood  of the  statistics implied by values $\boldsymbol\mu_\text{J}$, i.e.\ the probability mass (when sampling from $\mathfrak{u}$) of inner \totem{plex} given the ambient statistics $\boldsymbol\mu_\text{I}$. By definition of conditional sampling in the ambient \totem{plex} (see in particular Equation~\ref{eq:MultinomialSampling:PartitionFunction}) and the Pythagorean Lemma, the test statistic $t=2 N \infdiv{\mathfrak q_\text{J}}{\mathfrak q_\text{I}}$ of Equation~\ref{eq:iTest:ChiSquaredStatistics} tells us how the ratio of probability masses $\text{Pr}(\mathcal T_\text{J}) / \text{Pr}(\mathcal T_\text{I})$ distributes at large sample size.
As such, it is the data-centric analog to the model-centric likelihood ratio test. 

As a final remark on the emergence of spherical symmetry, we emphasize that this arises at large $N$ in $p_x$ space only, when considering  distributions with a given $I$-divergence, i.e.\ a given sampling probability under the reference distribution. Since 
\begin{equation}
    \mathfrak p\in\mathcal T\,:\quad\infdiv{\mathfrak p}{\mathfrak q} = \infdiv{\mathfrak p}{\mathfrak u}  - \boldsymbol\theta \cdot\boldsymbol\mu \quad\text{with}\quad\mathfrak q = \argmin_{\mathfrak p\in\mathcal T}\infdiv{\mathfrak p}{\mathfrak u} \quad\text{and}\quad\boldsymbol\theta\equiv\boldsymbol\theta(\boldsymbol\mu)~,
\end{equation}
the test statistic of Equation~\ref{eq:ChiSquaredStatistics} and Equation~\ref{eq:iTest:ChiSquaredStatistics} will generically be a complicated, non-linear function of the values $\mu_a$ in attribute space. Therefore, there is no general spherical symmetry (hence no $\chi^2$ distribution) arising in the space of attribute manifestations $\underaccent{\bar}{x}\in\mathcal M$. Although \totem does not directly dictate how the random variables themselves are distributed, it reveals the universal distribution of the probabilities of their joint manifestations. 

\subsection*{Supplementary Text - Application}
To apply the formalism developed in the preceding paragraphs, we focus on problems with two random variables, the generalization to more attributes being straight-forward. 

\paragraph{Deriving hypotheses from power laws}
A purported fundamental law pinpoints a relationship $g(X,Y)$ between attributes $X$ and $Y$, with any observed deviation from the law being solely associated with measurement errors or some known source of systematic distortion. In other words, whenever observed manifestation $(x,y)$ does not verify $g(x,y)$, there is a very clear, microscopic explanation for that violation.  
In contrast, an effective law predicts only some expected relationship $\langle g(X,Y)\rangle$ between the observable attributes allowing for multiple sources of microscopic violation, i.e.\ individual manifestations do not need to obey $g$.
Given some complex theory which is perceived as fundamental, we can often derive or at least qualitatively justify expected relationships that are hopefully amenable to experimental verification.

When lacking an understanding of the underlying processes that generate data or when the study fails to systematically control for them, we usually refer to any suspected effective law as empirical. In the \totem description, we propose the form of hypothesis condition associated to an empirical law.  
After constructing the manifestation distribution that fulfills the hypothesis condition, we apply the universal $I$-test to assess the observed violation of this condition in the data.

For the purpose of allometric relationships, we regard the body mass of a species as the independent (explanatory) variable $X$ and investigate the basal metabolic rate (\textsc{bmr}) as the response variable $Y$.
An effective law --\,as opposed to a more fundamental one\,-- allows for violations as long as its form is obeyed in expectation:
\begin{equation}
\label{eq:PowerLaw:expectation}
\langle Y \rangle = a^{(\text b)} \langle X^\text{b}\rangle ~,
\end{equation}
for given $\text{b}\in(0,1)$. 
As long as the suspected relationship stays empirical, $a^{(\text b)}\in\mathbb R_+$ cannot be predicted by some more fundamental theory and remains a study-dependent proportionality factor. This gives rise to a family of law-driven hypotheses parameterized by $a^{(\text b)}$. Naively fixing this factor from the data by substituting expectations with their empirical estimates does not constitute a test of the law, but a mere restating of what we have already known: the empirical distribution $\mathfrak f$ is most likely under itself (see Equation~\ref{eq:MutlinomialDistro} with $\mathfrak u=\mathfrak f$) when $a^{(\text b)}=\langle Y \rangle_{\mathfrak f}/\langle X^\text{b}\rangle_{\mathfrak f}$ is satisfied. 

To circumvent our theoretical ignorance about the proportionality factor, we define a random variable 
\begin{equation}
A^{(\text b)} = \frac{Y}{X^\text{b}}
\end{equation}
which --\, seen as a field $a^{(\text b)}(x,y)\in\mathbb R_+$\,-- inherits the randomness from the response $Y$.
Subsequently, we require absence of correlation between  fluctuations of $X^\text{b}$ and $A^{(\text b)}$, i.e.\ vanishing covariance\footnote{This correlator is invariant under scale transformation $X\rightarrow c X$, as it should.}
\begin{equation}
\label{eq:PowerLaw:VanishingCorrelator}
\left\langle(A^{(\text b)}-\langle A^{(\text b)}\rangle)(X^\text{b}-\langle X^\text{b}\rangle)\right\rangle 
\overset{!}{=}0
\end{equation}
which in turn implies 
\begin{equation}
\label{eq:PowerLaw:EffectiveForm}
\langle Y\rangle = \left\langle \frac{Y}{X^\text{b}}\right\rangle\langle X^\text{b}\rangle~,
\end{equation}
where all expectations are taken over some ideal population. Contingent on the study design, we could either prespecify $\langle X^\text{b}\rangle$ in an experimental study or optimize it in a fully observational study. 

In the language of ordinary regression analysis, the deduced form of the law in Equation~\ref{eq:PowerLaw:EffectiveForm} constitutes the most efficient hypothesis in the family of Equation~\ref{eq:PowerLaw:expectation}. To see this define the mean squared error of random variable $A^{(\text b)}$ from some anticipated $a>0$,
\begin{equation}
MSE = \left\langle\left(\frac{Y}{X^\text{b}} - a\right)^2\right\rangle~,
\end{equation}
which is clearly minimized by 
\begin{equation}
a^* = \left\langle\frac{Y}{X^\text{b}}\right\rangle = \langle A^{(\text b)} \rangle~.
\end{equation}
Therefore, the condition of vanishing correlation in Equation~\ref{eq:PowerLaw:VanishingCorrelator} selects the hypothesis that automatically minimizes the $MSE$ of the proportionality factor. 

In \totem, expectations $\langle\cdot\rangle:=\langle\cdot\rangle_{\mathfrak p}$ are calculated using distributions $\mathfrak p$ from the simplex 
\begin{equation}
\label{eq:XY_Simplex}
\mathcal P = \left\{ p(x,y)\in\mathbb R_{\geq0} ~\vert~  \sum_{x,y} p(x,y)=1\right\}
\end{equation}
over recorded joint manifestations $(x,y)$  of $X$ and $Y$. In our phenomenological analysis, a good interspecific law is not expected to exclude any species by setting  $p(x,y)=0$ (unless there is some crude error in the corresponding data point). Hence, we anticipate that all observed manifestations $(x,y)\in\mathcal M$ admit non-vanishing probabilities under the hypothesis of the law. If some probability were forced to zero on account of the hypothesis conditions, the law would be usually rejected ($p\text{-value}\rightarrow0$).

With prespecification of mass in an experimental study, we arrive at the composite  hypothesis 
\begin{equation}
\label{eq:TOTEM:ExperimentalStudy:hypothesis_structural_conditions}
\langle X^\text{b}\rangle_\mathfrak{p} = \langle X^\text{b}\rangle_\mathfrak{f} \quad\text{and}\quad
\left\langle Y -  \langle X^\text{b}\rangle_\mathfrak{f}\: \frac{Y}{X^\text{b}} \right\rangle_\mathfrak{p} = 0~.
\end{equation}
This consists of structural and hypothesis conditions, respectively. To write expectations we have assumed no concrete error distribution, as historically done in the literature~\cite{seal_regression_1967}. We refer to the \totem{plex} whose distributions fulfill the composite hypothesis as the hypothesis \totem{plex} $\mathcal T_\text{H}$. Within the latter, $\infdiv{\mathfrak p}{\mathfrak f}$ is minimized to obtain the hypothesis distribution $\mathfrak p_\text{H}\in\mathcal T_\text{H}$. 

In an observational study, moments are usually not prespecifiable. Writing Equation~\ref{eq:TOTEM:ExperimentalStudy:hypothesis_structural_conditions} as  
\begin{equation}
\label{eq:TOTEM:observationalStudy:hypothesis_conditions}
\langle X^\text{b}\rangle_\mathfrak{p} = \mu_X \quad\text{and}\quad
\left\langle Y -  \mu_X\: \frac{Y}{X^\text{b}} \right\rangle_\mathfrak{p} = 0~,
\end{equation}
we thus have to contemplate all empirically admissible values of $\mu_X>0$. In such a design, the hypothesis \totem{plex} includes all distributions satisfying the effective power law for some value of the fractional moment of $X$. 
For each possible $\mu_X$ we compute $\mathfrak q$, the $I$-projection of $\mathfrak f$ on $\mathcal T_\text{H}\equiv\mathcal T_\text{H}(\mu_X)$. Eventually, we are interested in the value $\mu_X^*$ which allows for the absolute minimum of the $I$-divergence in the given problem:   
\begin{equation}
\label{eq:XY:ObservationalStudy:DoubleMinimization}
\mu_X^*=\argmin_{\mu_X} \infdiv{\mathfrak q}{\mathfrak f} \quad\text{with}\quad \mathfrak q = \argmin_{\mathfrak p\in\mathcal T_\text{H}(\mu_X)} \infdiv{\mathfrak p}{\mathfrak f} ~.
\end{equation}
As long as we consider all possible values for the fractional moment, our double minimization procedure amounts to minimizing the $I$-divergence from $\mathfrak f$ under the combined  condition 
\begin{equation}
\langle Y \rangle_\mathfrak{p} = \langle X^\text{b}\rangle_\mathfrak{p} \left\langle\frac{Y}{X^\text{b}} \right\rangle_\mathfrak{p} 
~,
\end{equation}
which is non-linear in the probabilities $p(x,y)$. Albeit any valid parametrization of this non-linear conditional minimization of the $I$-divergence is formally equivalent, e.g.\ in terms of the expected ratio $\mu_A=\left\langle\frac{Y}{X^\text{b}} \right\rangle_\mathfrak{p}$ instead of $\mu_X=\langle X^\text{b}\rangle_\mathfrak{p}$, they could differ in their numerical stability. 

\paragraph{Application of the $I$-test}
Now, we are in a position to test the competing empirical laws for $\text{b}\in\{2/3,3/4\}$.
The scaling exponent $2/3$ is supported by Rubner's surface law~\cite{rubner_uber_1883} that explains metabolic scaling with the heat loss over the animal's surface. 
The exponent $3/4$ is supported by the West, Brown and Enquist model (WBE) linking metabolic scaling to a fractal nutrient-supply network, as outlined in the main text.
A variable, species- or taxon-specific exponent $b\in[0,1]$ would mean that there is no effective law to test for~\cite{agutter_metabolic_2004} and is not further considered here.
 
First, we  need to construct the hypothesis distribution under the hypothesis and structural conditions. 
In the linear algebraic formulation of empiricism, normalization on the probability simplex,  
\begin{equation}
\label{eq:XY:structuralconstraints:normalization}
g_\text{N}(x,y) = 1 ~,
\end{equation}
is always implied as a structural condition on probabilities. 
A finite number of species on earth implies a finite number of unique $(x,y)$ tuples, leading to a well-defined indexing of the column space of $\mathbf G$ in Equation~\ref{eq:LinearSystem}. 
The effective law itself corresponds to a row in the coefficient matrix of the form 
\begin{equation}
g_\text{H}(x,y) = y - \mu_X \frac{y}{x^\text b}   ~.
\end{equation}
Depending on the study design, we supplement the deduced hypothesis condition by further structural condition(s). If body mass were prespecified in an experimental study, we need the additional row 
\begin{equation}
\label{eq:Xb:structuralconstraint}
g_X(x,y) = x^{\text b} ~,
\end{equation}
so that conditions~\ref{eq:TOTEM:ExperimentalStudy:hypothesis_structural_conditions} are represented as
\begin{equation}
\label{eq:ExperimentalStusy:LinearSystem}
\mathbf G = 
\renewcommand{\arraystretch}{0.9}
\begin{pmatrix} \vec g_\text{N}\\  \vec g_X\\ \vec g_\text{H}  
\end{pmatrix} 
\quad\text{and}\quad
\boldsymbol\mu_\text{H} =
\begin{pmatrix}
1\\
\langle X^\text{b} \rangle_{\mathfrak f}\\
0
\end{pmatrix} 
~.
\end{equation}

With the linear system at hand, the hypothesis \totem{plex} is defined on the simplex $\mathcal P$ from Equation~\ref{eq:XY_Simplex} by 
\begin{equation}
\mathcal T_\text{H}  = \left\{\mathfrak p\in\mathcal P ~\vert~  \mathbf G\: \mathfrak p = \boldsymbol \mu_\text{H} \right\}~.
\end{equation}
Generically, $\mathfrak f\not\in\mathcal T_\text{H}$ unless there is some special symmetry which forces the law to be individually obeyed in absence of detectable measurement errors (making it thus a fundamental law of nature).
The $I$-projection of $\mathfrak f$ on $\mathcal T_\text{H}$ is the least-bias --\,in the information-theoretic sense\,-- distribution 
\begin{equation}
\label{eq:XY:null_iProjection}
\mathfrak p_\text{H} = \argmin_{\mathfrak p\in\mathcal T_\text{H}} \infdiv{\mathfrak p}{\mathfrak f}
\end{equation}
that implements the hypothesis alongside the structural conditions in Equation~\ref{eq:ExperimentalStusy:LinearSystem}, while staying closest to the data. Operationally, this happens as an optimization program that ``softly'' fixes the $k=\vert\mathcal M\vert-D$ ($D=3$ in this study) unconstrained directions, as explained in the paragraph about the $I$-projection.
To return from $\mathfrak p_\text{H}\in\mathcal T_\text{H}$ back to the effects measured in the data, we analogously define an alternative \totem{plex}
\begin{equation}
\label{eq:XY:alternativeTOTEMplex}
\mathcal T_\text{A}  = \left\{\mathfrak p\in\mathcal P ~\vert~ \mathbf G\: \mathfrak p = \boldsymbol \mu_\text{A} = \mathbf G\: \mathfrak f\right\}~,
\end{equation}
which by construction has at least one member, the empirical distribution itself.

For the \totem $\chi^2$ statistic, we need the $I$-projection of the hypothesis distribution on the alternative \totem{plex}. As long as the conjectured values $\boldsymbol \mu_\text{H}$ do not force some probability to vanish, $I$-projecting $\mathfrak p_\text{H}$ back onto $\mathcal T_\text{A}$ 
coincides with the empirical distribution,
\begin{equation}
\label{eq:XY:AlternativeDistribution_EmpiricalDistribution}
\mathfrak p_\text{A} =  \argmin_{\mathfrak p\in\mathcal T_\text{A}} \infdiv{\mathfrak p}{\mathfrak p_H} = \mathfrak f
~.
\end{equation}
To see this parameterize $\mathfrak p_\text{A}$ according to Equation~\ref{eq:iProj:parametricForm} in terms of Lagrange multipliers $\boldsymbol\theta\in\mathbb R^D$ such that all conditions in $\boldsymbol \mu_\text{H}$ are met, since by assumption hypothesized conditions do not force any vanishing probability.
Inspecting the $I$-divergence from the hypothesis distribution to the observed \totem{plex}, 
\begin{equation}
\infdiv{\mathfrak p}{\mathfrak p_\text{H}} = \infdiv{\mathfrak p}{\mathfrak f} - \boldsymbol \theta^T \mathbf G\: \mathfrak f
\quad{\text{for}}\quad \mathfrak p\in\mathcal T_\text{A}~,
\end{equation}
we see that the second term remains constant given the data, i.e.\ $\mathfrak f$ (note that $\boldsymbol\theta\equiv\boldsymbol\theta(\boldsymbol\mu_\text{H})$), while the first term attains its global minimum at  $\mathfrak f\in\mathcal T_A$. Hence, $\mathfrak p_\text{A}=\mathfrak f$. 

In case that some hypothesized condition(s) force at least one of the probabilities of observed manifestations to zero, we might fail to project back on $\mathcal T_\text{A}$ to recover $\mathfrak f$, implying a vanishing $p$-value (diverging $I$-divergence).
Indeed,  the construction of hypothesis distribution for $\text b=2/3$ requires the removal of the species \textit{Orcinus orca} (killer whale). 
Without \textit{Orcinus orca} the value of the observed violation of hypothesis condition cannot be achieved such that the probability of the data given Rubner's hypothesis is zero.

Notice that both hypothesis and alternative \totem{plices} are manifestly nested in the sense of Equation~\ref{eq:NestedTOTEMs:CoefficientMatrix} and Equation~\ref{eq:NestedTOTEMs:moments} within a structural \totem
\begin{equation}
\mathcal T_\text{S} = \left\{\mathfrak p\in\mathcal P ~\vert~ \vec g_\text{N}\cdot\left(\mathfrak p - \mathfrak f \right)= 0 \quad\text{and}\quad \vec g_X\cdot \left(\mathfrak p - \mathfrak f \right)= 0\right\}
\end{equation}
implementing structural conditions Equation~\ref{eq:XY:structuralconstraints:normalization} and Equation~\ref{eq:Xb:structuralconstraint}.
In particular, all distributions on $\mathcal T_\text{S}$ adhere to the mean of $X^\text{b}$ originally measured in the data assuming an experimental study design. For the formulation of the $I$-test in Equation~\ref{eq:iTest:ChiSquaredStatistics}, only the region of the simplex confined on $\mathcal T_\text{S}$ will be of interest. 
Starting from $\mathfrak p_\text{H}\in\mathcal T_\text{H}\subset\mathcal T_\text{S}$ as our reference distribution, the 
probability to sample a dataset with $\mathfrak p\in\mathcal T_\text{S}$ is dictated due to Equation~\ref{eq:ChiSquaredStatistics} and Equation~\ref{eq:TOTEMplex:ChiSquaredDistro} by the density
\begin{equation}
\label{eq:XY:concentrationTheorem}
\rho(\mathfrak p \,\vert\, \mathcal T_S) = \chi^2_{\vert{\mathcal M}\vert-2}\left(2 N \infdiv{\mathfrak p}{\mathfrak p_H} \right)~, 
\end{equation}
to leading order in the large $N$. The cardinality  of the set of manifestations $\vert\mathcal M\vert$ is calculated by the total number of distinct tuples $(x,y)$ recorded in the data. 

To apply Equation~\ref{eq:iTest:ChiSquaredStatistics} on this setting, we note that the structural \totem{plex} plays the role of the ambient ($\text{I}\rightarrow\text{S}$) and the alternative \totem{plex} of the inner ($\text{J}\rightarrow\text{A}$). Hence, the $N$-leading approximation to the conditional probability mass of $\mathcal T_\text{A}$ is given by 
\begin{equation}
\label{eq:XY:iTest:chiSquared}
\rho( \mathcal T_\text{A}\, \vert\, \mathcal T_\text{S}) = \chi^2_{k}\left(2 N \infdiv{\mathfrak f}{\mathfrak p_\text{H}} \right)
\quad\text{with}\quad k =
(\vert\mathcal M\vert - 2) - (\vert\mathcal M\vert - 3) = 1 ~,
\end{equation}
when sampling from hypothesis distribution which plays the role of reference distribution $\mathfrak u=\mathfrak p_\text{H}$. Since $\mathfrak p_\text{H}$ belongs to the structural \totem{plex} $\mathcal T_\text{S}$ in the current construction, the $I$-projection of $\mathfrak p_\text{H}$ on the ambient $\mathcal T_\text{S}$ is $\mathfrak p_\text{H}$ itself. As established in Equation~\ref{eq:XY:AlternativeDistribution_EmpiricalDistribution}, the $I$-projection of $\mathfrak p_\text{H}$ on the inner \totem{plex} $\mathcal T_\text{A}$ recovers back $\mathfrak f$ (in absence of zero probabilities). 
Within the structural \totem{plex}, Equation~\ref{eq:XY:iTest:chiSquared} reflects the sampling probability of all nested \totem{plices}  defined by violations of the effective law that are associated to an $I$-divergence from $\mathfrak p_\text{H}$ at least as large as the observed $\infdiv{\mathfrak f}{\mathfrak p_\text{H}}$. 

To this end, we identify the cumulative probability over the upper tail of  $\chi^2_{k}$ with the model-agnostic, data-centric
\begin{equation}
\label{eq:XY:iTest:p_value}
\text{$p$-value} = 1 - F_{\chi^2}(2\:N \infdiv{\mathfrak f}{\mathfrak p_H}; 1)
\end{equation}
to sample from the hypothesis distribution $\mathfrak p_\text{H}$ any distribution $\mathfrak p\in\mathcal T_\text{S}$ that violates the effective law at least as much as measured in the data via $\vec g_\text{H}\cdot \mathfrak f\neq 0$. 
Applying the $I$-test on the 549 mammalian species, we find a $p\text{-value}=0.001871$ for $\text b = 3/4$ and $p\text{-value}=0$ (corresponding to $\infdiv{\mathfrak f}{\mathfrak p_\text{H}}=\infty$) for $\text b = 2/3$ under an experimental study design in Equation~\ref{eq:TOTEM:ExperimentalStudy:hypothesis_structural_conditions}. 

Although the so-called ``lady tasting tea'' scenario   does not represent a sensible study design concerning allometric theories, we provide its \totem realization for the sake of completeness. In addition to Equation~\ref{eq:ExperimentalStusy:LinearSystem}, the outcome is fixed by  
\begin{equation}
\label{eq:Y:structuralconstraint}
g_Y(x,y) = y ~,
\end{equation}
so that we augment the linear system with another non-trivial structural condition:  
\begin{equation}
\renewcommand{\arraystretch}{0.9}
\mathbf G = \begin{pmatrix} \vec g_N\\ \vec g_X\\ \vec g_Y \\  \vec g_\text{H}  \end{pmatrix} 
\quad\text{and}\quad
\boldsymbol\mu_\text{H} =
\begin{pmatrix}
1\\
\langle X^\text{b} \rangle_{\mathfrak f}\\
\langle Y \rangle_{\mathfrak f}\\ 
0
\end{pmatrix} 
~.
\end{equation} 
This amount of certainty leads to an exceedingly small $p\text{-value}=\mathcal O(10^{-15})$ for $\text b = 3/4$  and  $p\text{-value}=0$ for $\text b = 2/3$, as before.
In the opposite regime of minimal prespecification in e.g.\ a fully observational study which is expected by design to show less statistical power, see Equation~\ref{eq:TOTEM:observationalStudy:hypothesis_conditions}, a higher $p\text{-value} = 0.01115$ can be computed for Kleiber's law, while Rubner's law still exhibits vanishing  $p\text{-value}$. 

\paragraph{The logic of ``old statistics'' in empiricism} 
In this paragraph, we explain the inadequacy of conventional approaches to capture the essence of  empirical laws and thus the inability of the ``old statistics'' to design an appropriate statistical test.  For that, we use the standard paradigm of regression analysis. 
In ordinary non-linear regression,  the null hypothesis for pure allometry can be formulated in terms of a conditional expectation of the dependent variable $Y$: 
\begin{equation}
\label{eqn::Regresion_in_Allometry}
    \frac{\langle Y \delta(X,x)\rangle}{\langle\delta(X,x)\rangle} = a^{(\text{b})} x^\text{b}~.
\end{equation}
Evaluated on a set of observed masses $\lbrace x_1,\ldots, x_N \rbrace$, $\delta(x,x_i)$ represents  Dirac's delta function. The expectation over $Y$ is understood over the ``true'' model of the population. The conditional rule implies that $y$-values are systematically distributed around the allometric prediction with an error $\varepsilon$ (estimated via the residuals $\varepsilon_i = y_i - a^{(\text{b})}x_i^\text{b}$), which is taken to be normally distributed, $\varepsilon \sim \mathcal N(0;\sigma)$. 

This logic greatly  differs from our definition of an empirical power law in Equation~\ref{eq:PowerLaw:expectation}.
Lacking a deeper, more {fundamental} understanding of the involved processes or because of their complexity, we accept that it is of little sense to make statements about all those effects we cannot account for. Hence, we impose a global version of condition Equation~\ref{eqn::Regresion_in_Allometry}, which can be seen as a necessary basis for any more fundamental law --\,if latter is at all applicable. In particular, our effective law allows for heteroscedasticity as well as violations of normality. In fact,  its form imposes no distributional constraints on the random variables. 

Evidently, the more fundamental regression rule in Equation~\ref{eqn::Regresion_in_Allometry} increases statistical power, as it tests for all the error-generating assumptions. As we have exemplified by considering various study designs in the previous paragraph, increased statistical power is easy to obtain in \totem. In that sense, we could easily apply the $I$-test for a stricter version of the allometric law, e.g.\ by a composite hypothesis over conditionals. It is our wish to stay maximally empirical that leads us to the effective form of the law in Equation~\ref{eq:PowerLaw:EffectiveForm}, which makes no statement about underlying processes in allometry that are beyond the theoretical understanding of the testable hypothesis and/or our experimental control.

In the premise of regression analysis, a cornerstone objective is the minimization of the mean squared error. Let us transcribe in the \totem framework this widely used paradigm in order to understand its information-theoretic implications. Albeit seemingly intuitive, objectives like the minimization of the dispersion of error $E = Y-a^{(\text b)}X^\text{b}$  result in unnecessary bias, conditioning the hypothesis distribution further away from $\mathfrak f$ than an effective law like Equation~\ref{eq:PowerLaw:expectation} would  actually  require.
For example, the mildest condition of vanishing mean error $\langle E\rangle=0$ with lowest variance $\partial_{a^{(\text b)}}\langle E^2\rangle=0$ dictates to fix the unknown proportionality factor according to the hypothesis
\begin{equation}
a^{(\text b)} = \frac{\langle Y\rangle}{\langle X^\text{b}\rangle} = 
\frac{\langle Y X^\text{b}\rangle}{\langle X^{2\text{b}}\rangle}
~.
\end{equation}
In an experimental study, this would require prespecification of two moments of $X$. Hence, the hypothesis \totem{plex} would be  defined by 
\begin{equation}
\langle X^\text{b}\rangle_\mathfrak{p} = \langle X^\text{b}\rangle_\mathfrak{f} \quad{,}\quad
\langle X^{2\text{b}}\rangle_\mathfrak{p} = \langle X^{2\text{b}}\rangle_\mathfrak{f} 
\quad\text{and}\quad
\left\langle \langle X^{2\text{b}}\rangle_\mathfrak{f} \: Y  -  \langle X^\text{b}\rangle_\mathfrak{f} \: Y X^\text{b} \right\rangle_\mathfrak{p} = 0~.
\end{equation}
Because of three non-trivial conditions, the resulting hypothesis distribution $\mathfrak p_\text{H}'$ would be more biased w.r.t.\ the data than the hypothesis distribution associated to Equation~\ref{eq:TOTEM:ExperimentalStudy:hypothesis_structural_conditions}, meaning $\infdiv{\mathfrak p_\text{H}'}{\mathfrak f} \geq \infdiv{\mathfrak p_\text{H}}{\mathfrak f}$.

If we worked with $\log$-transformed quantities in the spirit of Huxley's multiplicative error model~\cite{huxley_problems_1993} using variables $\log X$ and $\log A = \log Y - b \log X$, then the preceding reasoning  about the absence of correlation between random variable $\log A$ and independent variable $\log X$ implies 
\begin{equation}
\left\langle(\log A-\langle \log A\rangle)(\log X-\langle \log X\rangle)\right\rangle 
\overset{!}{=}0\quad\Rightarrow\quad 
b = \frac{\langle \log Y \log X\rangle - \langle \log Y\rangle \langle \log X\rangle }{\langle(\log X)^2\rangle - \langle\log X\rangle^2}~.
\end{equation}
Albeit familiar from the setting of linear regression, since
\begin{equation}
\left\langle(\log A-\langle \log A\rangle)(\log X-\langle \log X\rangle)\right\rangle 
=
-\tfrac12\frac{\partial}{\partial b}  \left\langle \left(\log A - \langle\log A\rangle\right)^2 \right\rangle
~,
\end{equation}
this condition on expectations does not reproduce the effective law. Moreover, it is not generically straight-forward to back-transform the expectations to the original non-linear formulation, due to Jensen's inequality. The same applies, whenever regressing $\log Y$ against $\log X$ given slope $\text b$ and fitting the intercept $\log a^{(\text b)}$. Consequently, the optimization problem in $\log$ space has little to do with the desired effective law in Equation~\ref{eq:PowerLaw:expectation}. 

The logic of the bootstrap is intimately related to the starting point of \totem: the data --\,described  by the empirical distribution $\mathfrak f$\,-- provides the most objective statistical model about the population. To sample datasets for the estimation of confidence intervals and $p$-values, the bootstrap uses a concrete ``null'' distribution. To deduce the latter, observed manifestations of attributes are transformed in ways that mostly remain unwarranted in realistic applications. For example, one often assumes that there is merely a systematic effect which uniformly shifts manifestations of an attribute away from some ``true'' mean. Under the assumption of such shift symmetry, the ``null'' distribution can be easily obtained by shifting back all observed manifestations and resampling from $\mathfrak f$ which is seen as a translated version of the ``null'' distribution. However, this assumption not only neglects the multitude of other effects that would have influenced sampling, alongside study design, but most alarmingly it could extrapolate to non-physical manifestations of the attributes.

\paragraph{Mixed categorical-metric and pure categorical settings}
To exemplify the adaptability of our framework, we derive the \totem analog of the $t$-test. In a mixed setting with categorical variable $S$ signifying group membership $s=a,b$ (stated for simplicity in the binary setting) and metric response $Y$, we outline the empirical test for group means without relying on the variance of $Y$.

Using the projector $\delta_s(S)$  on group $s$, the null hypothesis of the equality of means is written as 
\begin{equation}
\label{eq:iTest:equalityOfGroupMeans}
\frac{\langle Y \delta_a(S)\rangle}{\langle\delta_a(S)\rangle} \overset{!}{=} \frac{\langle Y \delta_b(S)\rangle}{\langle\delta_b(S)\rangle}~.
\end{equation}
Incidentally, this hypothesis condition can be seen as a consequence of vanishing covariance between response $Y$ and projector $\delta_s(S)$, cf.\ Equation~\ref{eq:PowerLaw:VanishingCorrelator},
\begin{equation}
    \label{eq:YS:vanishing_covariance}
    \left\langle \left(Y - \langle Y\rangle \right) \left( \delta_s(S) - \langle\delta_s(S)\rangle \right) \right\rangle \overset{!}{=}0 \quad\text{for}\quad s=a,b~,
\end{equation}
using that $\langle\delta_a(S)\rangle+\langle\delta_b(S)\rangle=1$ and $\langle Y\delta_a(S)\rangle+\langle Y\delta_b(S)\rangle=\langle Y\rangle$.

In the linear-algebraic language, this immediately translates into a condition vector in terms of Kronecker deltas
\begin{equation}
\label{eq:YS:HypothesisVector}
g_0(y,s) = y \left[\frac{\delta_{s, a}}{\mu_a} - \frac{\delta_{s, b}}{1-\mu_a}\right]~,
\end{equation}
supplemented by  
\begin{equation}
g_\text{N}(y,s) = 1 \quad\text{and}\quad g_a(y,s) = \delta_{s, a} ~,
\end{equation}
which implement the marginal condition $\vec g_a\cdot \mathfrak p=\mu_a$ on the ideal population --\,otherwise conditional means would be impossible to define in the first place. Without loss of generality, we have parameterized our composite hypothesis in terms of the prevalence $\mu_a$ of the $a$-th group, since $(1-\vec g_a\cdot\mathfrak p)$ gives the prevalence of the $b$-th group. The coefficient matrix and hypothesized values in Equation~\ref{eq:LinearSystem} read
\begin{equation}
\label{eq:YS:coefficientMatrix}
\renewcommand{\arraystretch}{0.8}
    \mathbf G = \begin{pmatrix}
    \vec g_\text{N}\\
    \vec g_a\\
    \vec g_0
    \end{pmatrix}
    \quad\text{and}\quad
    \boldsymbol\mu_\text{H}=\begin{pmatrix}1\\ \mu_a\\ 0\end{pmatrix}
    ~,
\end{equation}
respectively.

Next, we define the hypothesis \totem{plex} over joint manifestations $(y,s)$, i.e.\ recorded outcomes in both groups. In an ideal experimental study with prespecifiable group prevalence, $\mu_a = \vec g_a \cdot \mathfrak f$ which we learn from the empirical marginal. 
In particular, the structural \totem{plex} $\mathcal T_\text{S}=\{ \mathfrak p\in\mathcal P ~\vert~\vec g_\text{N}\cdot \mathfrak p=1 ~\text{ and }~ \vec g_a \cdot \left(\mathfrak p-\mathfrak f\right) = 0\}$ learns the prevalences of both groups from the data. The alternative \totem{plex} is formally given by Equation~\ref{eq:XY:alternativeTOTEMplex} using the coefficient matrix $\mathbf G$ from Equation~\ref{eq:YS:coefficientMatrix}. 
 
Readily, we can apply the $I$-test via Equation~\ref{eq:XY:iTest:p_value} to compute the $p$-value for the observed deviation $\vec g_0 \cdot \mathfrak f$ from the equality of group means. 
Notice how the structure of the $N$-leading approximation to conditional sampling distribution in Equation~\ref{eq:XY:iTest:chiSquared}, being exclusively expressed in terms of $p_x$'s, does not formally change whether we test the effective law or group-mean equality.  
Crucially, there are no additional assumptions necessary to be able to apply the $I$-test.

If researchers had prespecified the group prevalences, anticipating a concrete value $\mu_a$ which however differed from the observed prevalence $\vec g_a\cdot\mathfrak f$, an additional uncertainty must be taken into account. Fortunately, the $I$-test remains flexible enough for that. In fact, we proceed as before using the linear system in Equation~\ref{eq:YS:coefficientMatrix} to construct the hypothesis \totem{plex} where $\mu_a$ has been prespecified. This time the marginal $\vec g_a \cdot \mathfrak p$ differs between $\mathcal T_\text{H}$ and $\mathcal T_\text{A}$ so that the structural \totem{plex} coincides with the simplex. The \totem $\chi^2$ statistic has $k=2$ degrees of freedom, signifying the decrease in statistical power due to lack of control in the group prevalence.
  
Last but not least, we consider a binary response $Z$ with outcomes $z\in\{\texttt{success},\texttt{failure}\}$ distributed over the two groups $s=a,b$ from above. The null hypothesis of equal rates in both groups,
\begin{equation}
    \frac{\langle \delta_\texttt{success}(Z) \delta_a(S) \rangle}{\langle \delta_a(S) \rangle} \overset{!}{=} \frac{\langle \delta_\texttt{success}(Z) \delta_b(S) \rangle}{\langle \delta_b(S) \rangle} ~,
\end{equation}
(cf.\ Barnard's test in experimental study or Pearson's test in observational study) translates into the hypothesis condition 
\begin{equation}
    \label{eq:ZS:HypothesisVector}
    g_0(z,s) = \delta_{z, \texttt{success}}\left[\frac{\delta_{s, a}}{\mu_a} - \frac{\delta_{s, b}}{1-\mu_a}\right]~.
\end{equation}
Incidentally, this is the condition vector of Equation~\ref{eq:YS:HypothesisVector}, if $Y$ were constrained to $y\in\{0,1\}$ after mapping the outcomes of $Z$, accordingly. Due to gauge symmetry in categorical variables, the condition for $z=\texttt{failure}$ is implied by normalization, see similar comment below Equation~\ref{eq:YS:vanishing_covariance}. Hence, the coefficient matrix is again given by Equation~\ref{eq:YS:coefficientMatrix} using the hypothesis vector in Equation~\ref{eq:ZS:HypothesisVector} and the $I$-test proceeds accordingly. 
Under an observational study design, we have to optimize the $I$-divergence in the prevalence $\mu_a\in(0,1)$ --\,at potential cost of statistical power, see Equation~\ref{eq:XY:ObservationalStudy:DoubleMinimization} thereof.
\newpage 


\begin{figure}[h] 
	\centering
	\includegraphics[width=1\textwidth]{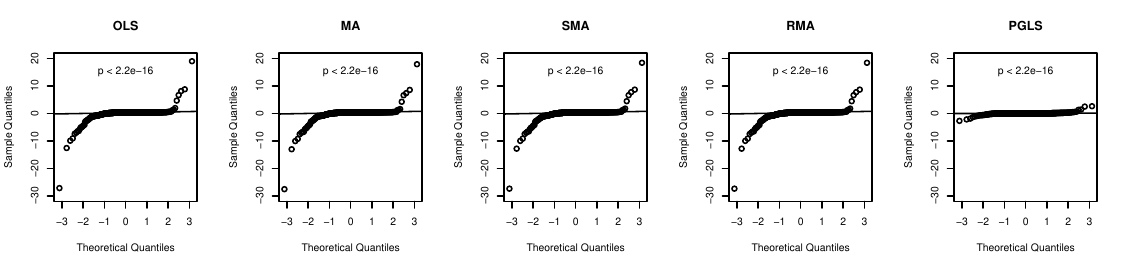} 

	\caption{\textbf{Normal QQ Plot of the Residuals of the fitted Model.}
	    Shown are the theoretical quantiles ($x$-axis) versus the quantiles of the residuals ($y$-axis  in $\ell~\text{O}_2~\text{h}^{-1}$) for the fitted models using ordinary least squares- (OLS), major axis- (MA), standardized major axis- (SMA), reduced major axis- (RMA), and phylogenetic generalized least squares (PGLS)-regression.  The almost horizontal line in each plot indicates the quantiles expected for the normal distribution. For each fitting method the $p$-value of the Shapiro-Wilk test is indicated, which is in all cases smaller than the machine precision of 2.2e-16.}
	\label{fig:figureS1} 
\end{figure}










\end{document}